\documentclass[
 aip,
 amsmath,amssymb,
preprint,%
 reprint,%
]{revtex4-1}

\usepackage{graphicx}
\usepackage{amsfonts}
\usepackage{amssymb}
\usepackage{amsmath}
\usepackage{ntheorem}
\usepackage{mathrsfs}
\usepackage{epsfig}
\usepackage{color}
\usepackage{algorithm}
\usepackage{algorithmic}

\usepackage[caption=false]{subfig}
\usepackage{pgfplots}
\pgfplotsset{compat=1.18}

\theoremheaderfont{\bfseries}
\theorembodyfont{\rm}

\newtheorem{teo}{Theorem}

\newtheorem{lem}{Lemma}

\newtheorem{prp}{Proposition}

\newtheorem{hyp}{Hypothesis}

\newtheorem{dfn}{Definition}

\theorembodyfont{\rm}

\newtheorem*{remn}{Remark}
\newtheorem{exa}{Example}

\newtheorem*{proofn}{Proof}

\newcommand{\vb}[1]{\textbf{#1}}

\DeclareMathOperator{\Arg}{Arg}

\newcommand{\iu}{\mathrm{i}}
\newcommand{\e}{\mathrm{e}}

\renewcommand{\S}{\mathbb{S}^1}

\newcommand{\ga}[1]{\textcolor{black}{{#1}}}

\usepackage{graphicx}
\usepackage{dcolumn}
\usepackage{bm}

\usepackage[utf8]{inputenc}
\usepackage[T1]{fontenc}
\usepackage{mathptmx}
\usepackage{etoolbox}

\makeatletter
\def\@email#1#2{%
 \endgroup
 \patchcmd{\titleblock@produce}
  {\frontmatter@RRAPformat}
  {\frontmatter@RRAPformat{\produce@RRAP{*#1\href{mailto:#2}{#2}}}\frontmatter@RRAPformat}
  {}{}
}%
\makeatother
\begin{document}

\newenvironment{psm}
{\left( \begin{smallmatrix}}
		{\end{smallmatrix} \right) }


%



\title[Towards the complete description of stationary states]{Towards the complete description of stationary states of a Bose-Einstein condensate in a one-dimensional quasiperiodic lattice: A coding approach}
\author{G. L. Alfimov\textsuperscript{*}}
\affiliation{ 
Moscow Institute of Electronic Engineering,
    Zelenograd, Moscow, 124498, Russia.}
\email{galfimov@yahoo.com.}
\author{A. P. Fedotov}%
\affiliation{ 
Moscow Institute of Electronic Engineering,
    Zelenograd, Moscow, 124498, Russia.
}

\author{Ya. A. Murenkov}
\affiliation{ 
Moscow Institute of Electronic Engineering,
    Zelenograd, Moscow, 124498, Russia.
}
\author{D. A. Zezyulin}
\affiliation{School of Physics and Engineering, ITMO University, St. Petersburg 197101, Russia
}
\date{\today}

\begin{abstract}
    We consider stationary states of an effectively one-dimensional Bose-Einstein condensate  in a quasiperiodic lattice.   We formulate sufficient conditions for a  one-to-one correspondence between the stationary states with a fixed chemical potential and the set of bi-infinite sequences over a finite alphabet. These conditions  can be checked numerically. A bi-infinite sequence can be interpreted as a code of the corresponding solution.   A numerical example demonstrates the coding approach using an alphabet of three symbols.
\end{abstract}

\maketitle

\begin{quotation}
Quasiperiodic systems, characterized by the interplay between multiple incommensurate frequencies, represent a fundamental class of models in modern physics. They   occupy an intermediate position between perfectly ordered periodic lattices and fully disordered systems, exhibiting a unique  blend of properties from both extremes. The complex properties of quasiperiodic lattices are already apparent in the linear regime, where their spectra exhibit Anderson localization, mobility edges, and fractal structure. The interplay between quasiperiodicity and nonlinearity introduces an additional layer of complexity. Atomic Bose-Einstein condensates (BECs) represent a perfect platform, where the combination of both effects can be reached experimentally.  Within the mean-field approximation, the stationary states of  a BEC are described by the Gross-Pitaevskii equation (GPE).  In the GPE  with a periodic potential, each stationary state has a countable infinity of exact translational copies. This is not the case for quasiperiodic lattices, which lack translational symmetry; hence, each stationary state becomes unique. It may therefore seem unlikely that the full set of stationary states in a quasiperiodic potential admits a complete, unified description. In this paper we demonstrate that, in some cases, the stationary states emerging in a   quasiperiodic lattice can be  classified via a  coding approach that establishes a one-to-one correspondence with the set of bi-infinite sequences over a specific alphabet. Our approach adapts a method originally developed for periodic lattices. The core idea is to filter out physically irrelevant singular solutions. The remaining regular solutions form a scarce set which can, in some cases, be completely described using symbolic dynamics. We extend this approach to the  quasiperiodic setting and demonstrate that  for a specific range of governing parameters  the stationary  states can be encoded by bi-infinite sequences of symbols from a finite alphabet.
\end{quotation}

\section{Introduction}\label{Sect:Intro}

The interplay between incommensurate lattices is fundamental to the theory of quasicrystals \cite{QC01,QC02}, with wide applications in solid-state physics \cite{QC03,QC04}, photonics \cite{Photonics01,Photonics02,Photonics03,Zilber2021}, and the study of atomic  Bose-Einstein condensates \cite{Lye,Exp08,Reeves,Luschen,Modugno10,Deissler10}.  Quasiperiodic systems  can be regarded as intermediate between periodically ordered and fully disordered systems,   retaining certain features of both. In particular,  one-dimensional quasiperiodic lattices  can exhibit  Anderson localization 
\cite{Diener,M09,Adhikari2009,Biddle}
and have   mobility edges in the energy spectra \cite{Boers2007,LiLiS17,S-P19,S-P22}, which separate spatially localized and extended states.  On the other hand, spectra of quasiperiodic systems can sometimes  be described in terms of generalized Bloch functions \cite{Dinaburg,LL22}, a concept typically associated with periodic systems.

The challenges posed by  quasiperiodic systems are clearly evident even in the linear regime, as exemplified by the spectrum of  the one-dimensional Schr\"odinger operator ${ L}=-\frac{d^2}{dx^2}+V(x)$. If the potential $V(x)$ is periodic, the band-gap structure of the spectrum follows from the Bloch theorem (1929). However, the spectrum becomes much more complex for  a quasiperiodic potential $V(x)$  \cite{Dinaburg,LL22,MP84,FSW90,S90,E92, S82,L05,PF92}.
Even for the prototypical bichromatic potential
\begin{gather*}
    V(x)=A_1\cos (2x)+A_2\cos(2\theta x+\delta),
\end{gather*}
where irrational $\theta$  and real  $\delta \in [0, 2\pi)$ are fixed numbers,
the lower spectral region typically contains discrete eigenvalues   with localized eigenfunctions, while the overall structure of the spectrum  depends on the parameters of the potential and is, generically, fractal \cite{S-P19}.  

It is then natural to expect that the interplay between quasiperiodicity and nonlinearity leads to a further escalation in complexity. Bose-Einstein condensates  (BECs) provide an ideal platform for probing this  regime experimentally.  Localization of a BEC  arises  from the interplay between the mean-field nonlinearity and the optical lattice potential generated by multiple laser sources \cite{P06}. If the lattice is periodic, the localized excitations appear at  chemical potentials in the spectral gaps  (the so-called {\it gap solitons}  \cite{BK04,MO06,LOK03,AKS02,PSK04,JY10}). Quasiperiodic optical lattices have been created in several experiments  by superimposing   standing waves with incommensurate wavelengths  \cite{Lye,Exp08,Luschen,Deissler10,Reeves}.  In parallel, the theoretical studies predict rich and complex behavior  for nonlinear modes  in one-dimensional quasiperiodic potentials  \cite{MS06,PK05,Kominis2008,Huang19,PKZ22,ZA24,Konotop24}.


Apart from the trap potentials, the localization of BEC may be facilitated by nonlinear-lattice pseudopotentials  corresponding to a spatially periodic modulation of the coefficient in front of the nonlinear term in the respective Gross-Pitaevskii equation \cite{KMT11}. This structure can be created in a BEC by means of the Feshbach resonance controlled by magnetic or optical
fields \cite{Pseudo01,Pseudo02}. Even when both the trap potential and pseudopotential are periodic, their frequencies may be incommensurate \cite{Pseudo03,Pseudo04}, rendering the overall problem  quasiperiodic.

Within the mean-field approximation, the dynamics of a cigar-shaped BEC are described by the Gross-Pitaevskii equation. In its dimensionless form, this equation is:
\begin{gather}
        \iu \Psi_t=-\Psi_{xx}+V(x)\Psi+P(x) |\Psi|^2\Psi.\label{GPEq}
\end{gather}
Here $\Psi(t,x)$ is the  macroscopic wavefunction, $V(x)$ is the trap potential, and $P(x)$ is the pseudopotential. Stationary states of the BEC correspond  to solutions of the form
\begin{gather}
\Psi(t,x)=\e^{-\iu \mu t}u(x),\label{Anzats}
\end{gather}
where   $\mu$ is the chemical potential, and $u(x)$ is the stationary wavefunction.  In this paper, we assume that $u(x)$ is real-valued. This assumption does not lead to a loss of generality if $u(x)$  is  a solitary wave which  tends to zero  $x\to \infty$ and $x\to-\infty$. At the same time, our analysis is  also applicable to  real-valued stationary solutions that do not satisfy   zero boundary conditions at $x\to \pm \infty$. The  stationary wavefunction   $u(x)$  solves the equation
\begin{gather}
u_{xx}+(\mu-V(x))u+P(x) u^3
=0.\label{StatGPEq}
\end{gather}

In periodic lattices, every stationary state has a countable set of exact replicas that are equivalent under a translation. This is not the case for quasiperiodic lattices, which lack translational symmetry; hence, each stationary state is unique. It may therefore seem unlikely that the full set of stationary states in a periodic potential admits a complete, unified description. The main goal of this paper is to demonstrate that, in some cases, the real-valued stationary states emerging in a quasiperiodic lattice can be fully classified via a transparent coding procedure that establishes a one-to-one correspondence with the set of bi-infinite sequences over a specific alphabet. Our approach represents a deep extension of a method originally developed for the periodic version of Eq.~(\ref{StatGPEq}), where $V(x)$ and $P(x)$ were periodic functions with a common period $\ell$ \cite{AA13,AKZ17,AK16,LAM16,AL23,AL24}. The original method used  the fact that  for a sign-alternating or negative  $P(x)$   most  solutions of the Cauchy problem for Eq.~(\ref{StatGPEq}) tend  to infinity at some finite point on the  real axis \cite{AL15}. We refer to such solutions as {\it singular}. These singular solutions are physically irrelevant. The remaining   {\it regular} solutions   form a scarce set which in some cases can  be  fully described using symbolic dynamics. When  $V(x)$ and $P(x)$ are periodic with a common period $\ell$, one can analyze Eq.~(\ref{StatGPEq}) using the  Poincaré map, defined as follows:
\begin{gather}
	\mathcal{P} 
         \begin{pmatrix} u_0 \\ u_0' \end{pmatrix}
	= \begin{pmatrix} u_\ell \\ u_\ell' \end{pmatrix}.
\label{eq:poincare-map}
\end{gather}
Here $u_\ell = u(\ell)$, $u_\ell' = u_x(\ell)$, and $u(x)$ is a solution of the Cauchy problem for Eq.~(\ref{StatGPEq}) with the initial conditions $u(0) = u_0$, $u_x(0) = u_0'$. Due to the presence of singular solutions, the Poincar\'e map is defined not on the whole plane of initial data $(u,u')$ but on some domain $\mathscr{U}^+_\ell\subset \mathbb{R}^2$. If the  action of $\mathcal{P}$ on $\mathscr{U}^+_\ell$ is hyperbolic and   conjugate to a Smale horseshoe map,
then the regular solutions can be described, either completely  or partially, by bi-infinite sequences of symbols (also called codes) from a finite alphabet \cite{AL23,AL24}. Verifying the hyperbolic properties of the map $\mathcal{P}$  requires checking the hypotheses formulated in Ref.~\cite{AL24}, either numerically or analytically.

This approach cannot be applied directly if $V(x)$ or $P(x)$ is quasiperiodic, or if both are periodic with incommensurate periods.  In the present study we develop an   extension of  this method to the quasiperiodic case. The extended approach covers various important prototypical cases, as follows
\begin{align}
   (a)~ & V(x)=A_1\cos x+A_2\cos (\theta x+\delta),\quad P(x)=-1,\label{Eq:CaseA}\\[2mm]
   (b)~& V(x)=A_1\cos x,\quad P(x)=-1+A_2\cos (\theta x+\delta),\label{Eq:CaseB}\\[2mm]
   (c)~& V(x)=0,\quad P(x)=-1+A_1\cos x+A_2\cos (\theta x+\delta),\label{Eq:CaseC}
\end{align}
where $A_{1,2}$ are real amplitudes, $0\leq\delta<2\pi$ and $\theta>1$ is an irrational number. 

The paper is organized as follows. Section~ \ref{Sec:Statement} introduces the main concepts and basic definitions used throughout the paper. In Sec.~ \ref{sec:SymbDyn} we formulate two conditions  that guarantee the existence of a one-to-one correspondence between regular solutions of quasiperiodic Eq.~(\ref{StatGPEq}) and bi-infinite sequences over  some alphabet. Direct verification of these conditions, however, is challenging. Therefore, in Sec.~\ref{sec:StripMap}, we formulate two theorems that render these conditions amenable to a numerical check. In Sec.~\ref{sec:Examples} we apply our approach to Eq.~(\ref{StatGPEq}) with quasiperiodic potential $V(x)$ with $P(x)\equiv -1$ (the case (a), Eq.~(\ref{Eq:CaseA})). Section~\ref{Sec:Concl} provides a  summary and discussion.


\section{The basic equation and definitions}\label{Sec:Statement}

\subsection{The basic equation}

In this study, we focus on the equation
\begin{gather}
    \ddot{u} =G_\delta(x,u),
     \label{eq:stationary_delta}
\end{gather}
where $\ddot{u}$ is the second derivative with respect to $x$, 
\begin{gather}
G_\delta(x,u)=F(x,\theta x+\delta,u),\quad \theta>1, \label{Eq:DefG_d}
\end{gather}
and   $0\leq\delta<2\pi$ is a { fixed given} parameter. It is assumed that the function $F(X,Y,u)$ satisfies the following  conditions:
\begin{itemize}
    \item[(A)]  $F(X,Y,u)\in C^1({\mathbb R}^3)$; 

    \item[(B)] $F(X,Y,0)=0$ for any $X,Y\in\mathbb{R}$;

    \item[(C)] $F(X,Y,u)$ is $2\pi$-periodic with respect to both $X$ and $Y$.
\end{itemize}

If $\theta$ is a rational number, the right-hand side of Eq.~(\ref{eq:stationary_delta}) is periodic with respect to $x$.   Without loss of generality, it can be assumed that the period is equal to $2\pi$
(this can be achieved by renormalizing the independent variable and rescaling the function $F(X,Y,u)$ itself).
Equation (\ref{StatGPEq}) with (pseudo)potentials given by either (\ref{Eq:CaseA}) or (\ref{Eq:CaseB}) can be cast in the form (\ref{eq:stationary_delta}) if we assume
\begin{align*}
     (a)\quad & F(X,Y,u)=-(\mu-A_1\cos X-A_2\cos Y)u+ u^3,\\
     (b)\quad & F(X,Y,u)=-(\mu-A_1\cos X)u+(1-A_2\cos Y)u^3,\\
     (c)\quad & F(X,Y,u)=-\mu u+(1-A_1\cos X-A_2\cos Y)u^3.
\end{align*}

\subsection{The space $\S$ and the auxiliary family of equations}

The quotient space $\mathbb{S}^1 = \mathbb{R} / 2\pi \mathbb{Z}$  is   a set of real numbers, where any two points $x, y$ are considered equivalent if their difference is an integer multiple of $2 \pi$, i.e., $x\sim y$ iff $x-y = 2\pi n$, $n\in \mathbb{Z}$. If $\Delta \in \S$, then without loss of generality it can be assumed that  $\Delta\in [0, 2\pi)$. From the geometric point of view, $\S$ can be identified with the unit circle parameterized by the rotation   angle $\Delta$. The sum (to be denoted as $\oplus$) and difference ($\ominus$) of any two elements from $\mathbb{S}^1$ will be  computed modulo $2\pi$.  Using the argument function from the complex analysis, for any $\Delta_{1,2} \in \S$ these operations can be defined as
\begin{equation}
\label{eq:oplus}
\Delta_1 \oplus \Delta_2 = \mathrm{Arg\,} \e^{\iu(\Delta_1 + \Delta_2)}, \qquad  \Delta_1 \ominus \Delta_2 =\mathrm{Arg\,}  \e^{\iu(\Delta_1 - \Delta_2)}, 
\end{equation}
where $+$ and $-$   denote the usual sum and difference in $\mathbb{R}$,  $\iu$ is the imaginary unit, $\mathrm{Arg }$ is the principal value of the argument, and we use the convention where $\Arg z \in [0, 2\pi)$. 

Equation~(\ref{eq:stationary_delta}) is  a representative   of {a family of equations}
\begin{gather}
    \ddot{u} =G_\Delta(x,u), \qquad G_\Delta(x,u)=F(x,\theta x+\Delta,u),
     \label{eq:stationary_Delta}
\end{gather}
parametrized by   $\Delta \in  \S$.

\subsection{Regular and singular solutions}

\begin{dfn}\label{Def:sing_sol} 
A solution  $u(x)$  of Eq. (\ref{eq:stationary_Delta}) is called   \emph{singular} on  an interval $[a;b]$  if it   tends to infinity at some finite point $x_c\in[a;b]$  approaching  $x_c$ from the left, from the right or from both sides, i.e.,
\begin{gather*}
    \lim_{x\to x_c-0}|u(x)|=\infty\quad \mbox{or}\quad \lim_{x\to x_c+0}|u(x)|=\infty.
\end{gather*}      
\end{dfn}
Alternatively, we say that $u(x)$ \emph{collapses} at $x=x_c$.

\begin{dfn}\label{Def:reg_sol}
    A solution is called \emph{regular} on $[a,b]$ if it is not singular on $[a,b]$. A solution is called \emph{regular on $\mathbb{R}$} if it is not singular on $[a,b]$ for any $a$ and $b$ such that  $-\infty < a < b<\infty$.
\end{dfn} 

Singular solutions are abundant for Eq.~(\ref{eq:stationary_Delta}). Indeed,  singular solutions are known to exist \cite{AL15}  for Eq.~(\ref{StatGPEq}) with the (pseudo)potentials (\ref{Eq:CaseA}) and (\ref{Eq:CaseB}).  The following proposition provides  sufficient conditions for the existence of  singular solutions for Eq.~(\ref{StatGPEq}).
\begin{prp}
	Let $\Omega$ be a small enough neighbourhood of a point $x_c$, such that $P(x_c)<0$ and $V(x) \in   C^2(\Omega)$,  $P(x) \in C^4(\Omega)$.
	Then there exist two $C^1$-smooth one-parametric 
        families of solutions for equation (\ref{StatGPEq}) that are defined in $\Omega$ and collapsing at the point
        $x = x_c$ while approaching from the left, $x < x_c$. These two families are connected by  the symmetry $u \to -u$.	Each of these families can be parametrized by a free   variable 
        $C \in \mathbb{R}$.
\label{prop:singular-families}
\end{prp}

Proposition~\ref{prop:singular-families}   does not assume  periodicity or quasiperiodicity of  $V(x)$ and $P(x)$.


\subsection{The map $\mathcal{P}$}


Consider $\mathcal{L}=\S\times\mathbb{R}^2$, the space of all triples $(\Delta,u,u')$ with $\Delta \in \S$ and $(u,u')\in \mathbb{R}^2$.

     The map $\mathcal{P}\colon\mathcal{L}\to \mathcal{L}$ is defined as follows: $\mathcal{P} ({\bf p}_0)={\bf p}_1$, where ${\bf p}_0=(\Delta_0,u_0,u_0')$, ${\bf p}_1=({\Delta_1}, u_1, u_1')$, and
     \begin{itemize}
         \item $\Delta_1 =  \Delta_0 \oplus 2\theta\pi$;
         \item $u_1=u(2\pi)$, $u_1'=\dot{u}(2\pi)$ where $u(x)$ is the solution of the Cauchy problem for Eq.~ (\ref{eq:stationary_Delta}) with $\Delta=\Delta_0$ and the initial conditions $u(0)=u_0$, $\dot{u}(0)=u_0'$. 
     \end{itemize}   
 

The inverse map $\mathcal{P}^{-1}\colon\mathcal{L}\to \mathcal{L}$ acts as follows:
\begin{gather*}
    \mathcal{P}^{-1}( {\bf p}_0)={\bf p}_{-1},
\end{gather*}
where ${\bf p}_0=(\Delta_0,u_0,u_0')$, ${\bf p}_{-1}=(\Delta_{-1}, u_{-1}, u_{-1}')$,    $\Delta_{-1} = \Delta_0 \ominus 2\theta \pi$,   $u_{-1}=u(-2\pi)$, $u_{-1}'=\dot{u}(-2\pi)$, and $u(x)$ is the solution of the Cauchy problem for Eq.~(\ref{eq:stationary_Delta}) with $\Delta=\Delta_0$ and the  initial conditions $u(0)=u_0$, $\dot{u}(0)=u_0'$.   

\begin{dfn}\label{def:slice}
We call $\Delta$-{\em slice}  the plane $\mathcal{L}_{\Delta}\subset{\mathcal{L}}$ for a fixed $\Delta$, $\mathcal{L}_{\Delta}=\left.{\mathcal{L}}\right|_\Delta$.
\end{dfn}

\begin{remn}
    It follows from the definition of ${\mathcal P}$ that its restriction from $\mathcal{L}$ to $\mathcal{L}_\Delta$  acts to $\mathcal{L}_{\Delta\oplus 2\theta\pi}$, and the restriction of   ${\mathcal P}^{-1}$ to $\mathcal{L}_\Delta$  acts to  $\mathcal{L}_{\Delta\ominus 2\theta\pi}$, i.e.,
    \begin{gather*}
        {\mathcal P} \restriction _{\mathcal{L}_\Delta}\colon{\mathcal L}_\Delta\to {\mathcal L}_{\Delta\oplus 2\theta \pi},\quad {\mathcal P}^{-1}\restriction _{\mathcal{L}_\Delta}\colon{\mathcal L}_\Delta\to {\mathcal L}_{\Delta\ominus 2\theta \pi}.
    \end{gather*}

\end{remn}

\begin{dfn}\label{def:orbit}
    {\it An orbit} is a sequence of points $\{{\bf p}_n\}$, ${\bf p}_n\in \mathcal{L}$,   such  that $\mathcal{P}({\bf p}_n)={\bf p}_{n+1}$, $n=0,\pm 1,\ldots$.
\end{dfn}

Since  Eq.~(\ref{eq:stationary_Delta}) may have singular solutions,  the maps $\mathcal{P}$ and $\mathcal{P}^{-1}$ may not be  defined in the whole $\mathcal{L}$. Consequently, the iterations of $\mathcal{P}$ or  $\mathcal{P}^{-1}$ may stop after  a finite number of steps. Therefore, an orbit $\{{\bf p}_n\}$ is not necessarily  bi-infinite. The following  statement establishes a natural correspondence between the bi-infinite orbits of $\mathcal{P}$ and the regular solutions of Eq.~(\ref{eq:stationary_delta}).

\begin{lem}\label{Lem1} 
The solution $u(x)$ of the Cauchy problem for Eq.~(\ref{eq:stationary_delta}) with initial data $u(0)=u_0$, $\dot{u}(0)=u_0'$ is regular on $\mathbb{R}$ iff the orbit $\{{\bf p}_n\}$ generated by  $\mathcal{P}$ with ${\bf p}_0=(\delta,u_0,u_0')$  is bi-infinite. The regular solution $u(x)$ and the orbit $\{{\bf p}_n\}$ are related as   
\begin{gather*}
    {\bf p}_{n}= (\delta\oplus2n\theta \pi,u(2n\pi),\dot{u}(2n\pi)),\quad n\in\mathbb{Z}.
\end{gather*}
\end{lem}

\begin{proofn}
Let the solution $u(x)$  of Eq.~(\ref{eq:stationary_delta})  with $u(0)=u_0$, $\dot{u}(0)=u_0'$ exist   on $[0;2\pi]$. This is equivalent to  the map $\mathcal{P}$ being defined at ${\bf p}_0=(\delta,u_0,u_0')$ and, by definition of $\mathcal{P}$, we have   $\mathcal{P}({\bf p}_0)={\bf p}_1=(\Delta_1,u(2\pi),u'(2\pi))$, where $\Delta_1 = \delta\oplus 2\theta \pi$.

Let, in addition,  the same solution $u(x)$ exist    on $[2\pi, 4\pi]$. Due to $2\pi$-periodicity of $F(X,Y, u)$ in $X$ and $Y$,   this  is equivalent to  $\mathcal{P}$ being defined at ${\bf p}_1$, and by definition of $\mathcal{P}$, we have   $\mathcal{P}({\bf p}_1)={\bf p}_2=(\Delta_2,u(4\pi),u'(4\pi))$, where $\Delta_2 = \Delta_1 \oplus 2\theta \pi$. Using the definition of $\oplus$ in  Eq.~(\ref{eq:oplus}), it is easy to establish that $\Delta_2 = \delta \oplus 4\theta \pi$.

Repeating the procedure, we show that the existence of $u(x)$ on the semi-axis $x\geq 0$ is equivalent to the existence of  the infinite sequence ${\bf p}_0$, ${\bf p}_1$, ${\bf p}_2,\ldots$, ${\bf p}_{n+1}=\mathcal{P}({\bf p}_{n})$. Similarly,   $u(x)$ exists on the semi-axis $x\leq 0$  iff there exists the backward-infinite  sequence ${\bf p}_0$, ${\bf p}_{-1}$, ${\bf p}_{-2},\ldots$, where ${\bf p}_{n-1}=\mathcal{P}^{-1}({\bf p}_{n})$.   $\square$
\end{proofn}


Let us introduce several additional notations.
 Let $\mathscr{U}^+\subset \mathcal{L}$ be the domain of $\mathcal{P}$, and let  $\mathscr{U}^-\subset \mathcal{L}$ be the domain of $\mathcal{P}^{-1}$. Introduce 
\begin{gather*}
    \mathscr{U}=\mathscr{U}^+\cap\mathscr{U}^-.
\end{gather*}    
  For a fixed $\Delta_0 \in \S$ we denote $\mathscr{U}^+_{\Delta_0}= \mathscr{U}^+\cap \mathcal{L}_{\Delta_0}$, $\mathscr{U}^-_{\Delta_0}= \mathscr{U}^-\cap \mathcal{L}_{\Delta_0}$ and
\begin{gather*}
    \mathscr{U}_{\Delta_0}=\mathscr{U}^+_{\Delta_0}\cap\mathscr{U}^-_{\Delta_0}.
\end{gather*}  

\begin{remn}
The following comments are in order.

\begin{itemize}
  \item[(i)]   Note that ${\bf p}_0=(\Delta_0,u_0,u_0')\notin \mathscr{U}^+$ iff the solution of the Cauchy problem for the equation $\ddot{u} = G_{\Delta_0}(x,u)$
with the initial data $u(0)=u_0$, $\dot{u}(0)=u'_0$ collapses at some point $x=x_0 \in (0, 2\pi]$. Similarly, ${\bf p}_0=(\Delta_0,u_0,u_0')\notin \mathscr{U}^-$ iff the solution of the same Cauchy problem collapses at some point $x=x_0 \in [-2\pi, 0)$.  

\item[(ii)]   If $\theta=1$ then $\mathcal{P}(\mathscr{U}^+_{\Delta_0})\subset \mathcal{L}_{\Delta_0}$, $\mathcal{P}^{-1}(\mathscr{U}^-_{\Delta_0})\subset \mathcal{L}_{\Delta_0}$, for any $\Delta_0\in \S$.


\item[(iii)]   For any $\theta$ and $\Delta_0\in \S $ the following relations hold:
\begin{gather*}
 \mathcal{P}(\mathscr{U}^+_{\Delta_0})=\mathscr{U}^-_{\Delta_0\oplus  2\theta \pi},\quad  \mathcal{P}^{-1}(\mathscr{U}^-_{\Delta_0})=\mathscr{U}^+_{\Delta_0\ominus  2\theta \pi}.
\end{gather*}

\end{itemize}
\end{remn}

\subsection{Islands, Donuts, Curves and Strips}\label{sec:Islands}

\begin{dfn}\label{def:Lipschitz}
     Let $\gamma > 0$ be fixed. A function $f(x) \colon [a, b] \to \mathbb{R}$  is called  \emph{${\gamma}$-{Lipschitz} function} if for any 
     $x_1, x_2 \in [a, b]$ the inequality holds:
     \begin{equation}
		|f(x_1) - f(x_2)| \le \gamma |x_1 - x_2|.
     \end{equation}
\end{dfn}

\begin{remn}
    Evidently, if a function is a $\gamma_1$--{Lipschitz} function and $\gamma_2 > \gamma_1$	then $f(x)$ is also $\gamma_2$--{Lipschitz} function.
\end{remn}


\begin{remn}
    The union of $\gamma$-Lipschitz functions on $[a;b]$ for $\gamma\in[0;+\infty)$ constitutes the Banach space ${\rm Lip}(a;b)$ with respect to the norm \cite{W18}
    \begin{gather*}
        \|f(x)\|_{\rm Lip}=\sup_{x\in[a;b]}|f(x)|+\sup_{x,y \in[a;b],~x\ne y}\frac{\left|f(x)-f(y)\right|}{|x-y|}.
    \end{gather*}
    If a sequence $\{f_n(x)\}$ of $\gamma$-Lipschitz functions converges in ${\rm Lip}(a;b)$ to $f(x)$ then $f(x)$ is also a $\gamma$-Lipschitz function.  
\end{remn}


We call function $f(x)$ monotonically increasing if for $x_1 < x_2$, $f(x_1) \le f(x_2)$, and monotonically decreasing if $f(x_1) \ge f(x_2)$, i.e. the inequalities are not strict. We also say that for functions $f(x)$ and $g(x)$ \emph{the type of  monotonicity coincides} or $f(x)$ and $g(x)$ \emph{are of the same type of monotonicity} if both $f(x)$ and  $g(x)$ are increasing or both $f(x)$ and  $g(x)$ are decreasing functions simultaneously.

\begin{dfn}\label{def:Island}
A \emph{$\gamma$-island} is an open curvilinear quadrangle { $D_\Delta \subset \mathscr{U}_\Delta$} formed by two pairs of curves $\alpha^{\pm}$, $\beta^{\pm}$, such that:
\begin{itemize}
\item curves $\alpha^{\pm}$ do not intersect, they are graphs of monotonic $\gamma$-Lipschitz functions $u' = h_{\pm}(u)$ of the same type of monotonicity, and any solution of equation  \eqref{eq:stationary_Delta} with initial 
conditions $u(0)=u_0$, $\dot{u}(0)=u_0'$ such that $(u_0, u_0') \in \alpha^{\pm}$ collapses at the point $x = -2\pi$;
\item curves $\beta^{\pm}$ do not intersect, are graphs of monotonic $\gamma$-Lipschitz 
functions $u = v^{\pm}(u')$ of the same type of monotonicity, and any solution of equation \eqref{eq:stationary_Delta}
with initial conditions $u(0)=u_0$, $\dot{u}(0)=u_0'$ such that $(u_0, u_0') \in \beta^{\pm}$ collapses at the point
$x = 2\pi$;
\item if the functions $h^{\pm}(u)$ are increasing then $v^{\pm}(u')$ 
are decreasing, and vice versa, if functions $h^{\pm}(u)$ are decreasing then $v^{\pm}(u')$ are increasing respectively.
%
%
%
\end{itemize}	
\end{dfn}

\begin{remn}
    When defining $\gamma$-island we keep in mind the following. Let $\alpha^\pm$ be graphs of $\gamma_\alpha^\pm$-Lipschitz functions $u'=h_\pm(u)$ and $\beta^\pm$ be graphs of $\gamma_\beta^\pm$-Lipschitz functions $u=v_\pm(u')$. Taking
    \begin{gather*}
    \gamma=\max(\gamma_\alpha^-, \gamma_\alpha^+, \gamma_\beta^-,\gamma_\beta^+) 
    \end{gather*}
    one has that $u'=h_\pm(u)$ and $u=v_\pm(u')$ are $\gamma$-Lipschitz functions and they bound $\gamma$-island in the sense of Definition \ref{def:Island}.
\end{remn}

\begin{remn}
	If $D_\Delta$ is a $\gamma_1$-island and $\gamma_2 > \gamma_1$	then $D_\Delta$ also is a $\gamma_2$-island.
\end{remn}

\begin{remn}
        The $\gamma$-islands situated in $\mathcal{L}_\Delta$ can be regarded as connected components of the set
        $\mathscr{U}_\Delta$. At the same time we do not claim that in generic situation  the set $\mathscr{U}_\Delta$ is composed by $\gamma$-islands only.
\end{remn}
\begin{remn}
       The map $\mathcal{P}$ sends the island $D_\Delta\subseteq \mathscr{U}_\Delta\subset\mathcal{L}_\Delta$ to $\mathcal{L}_{\Delta_1}$ where $\Delta_1=\Delta\oplus  2\theta\pi$. Similarly, $\mathcal{P}^{-1}$ maps $D_\Delta$ to $\mathcal{L}_{\Delta_{-1}}$ where $\Delta_{-1}= \Delta\ominus  2\theta\pi$. The points at the boundaries $\beta^{\pm}$ of $D_\Delta$ are mapped by
       $\mathcal{P}$ to infinity at $\mathcal{L}_{\Delta_1}$. The points at $\alpha^\pm$  are mapped by 
       $\mathcal{P}^{-1}$  to infinity at $\mathcal{L}_{\Delta_{-1}}$.
\end{remn}
\begin{figure}
\centering
\includegraphics[width=\linewidth]{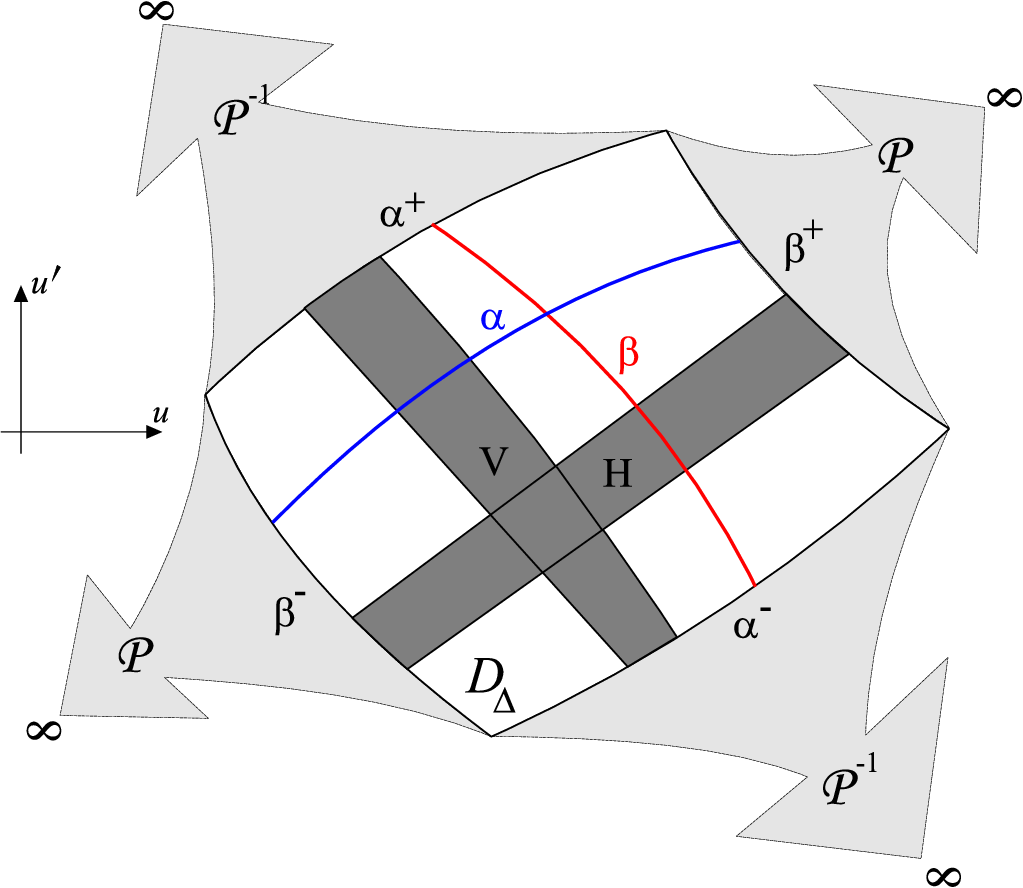}
\caption{An island $D_\Delta$ bounded by curves $\alpha^{\pm}$, $\beta^{\pm}$; $h$-curve $\alpha$, $v$-curve $\beta$, and two strips: $h$-strip $H$ and $v$-strip $V$.}
\label{fig:Island(def)}
\end{figure}
\begin{dfn}
    Let $D_\Delta\subset \mathcal{L}_\Delta$ be a $\gamma$-island bounded by curves $\alpha^{\pm}$, $\beta^{\pm}$. Let
    a curve $\alpha$  connect the opposite sides $\beta^{\pm}$ of the island $D_\Delta$. 
    \begin{itemize}
    \item We say that  $\alpha$  is \emph{${h}_{\gamma}$-curve} if it is a  graph of a monotonic $\gamma$-Lipschitz function $u' = h(u)$ and     its type  of monotonicity coincides with the functions $u' = h_{\pm}(u)$ that correspond  to the $\alpha^{\pm}$ boundaries of the island $D_\Delta$. 	
    \item We say that $\alpha$ is increasing/decreasing if it is a graph of \emph{increasing/decreasing} function  $u' = f(u)$.
    \end{itemize}
\label{def:h-curve}
\end{dfn}
\begin{dfn}
    We  call \emph{$h_{\gamma}$-strip} an open subset of $\gamma$-island $D_\Delta$ bounded by two $h_{\gamma}$-curves.
\label{def:h-strip}
\end{dfn}
\begin{dfn}
    Let $D_\Delta\subset \mathcal{L}_\Delta$ be a $\gamma$-island bounded by curves $\alpha^{\pm}$, $\beta^{\pm}$. Let a curve $\beta$ connect the opposite sides $\alpha^{\pm}$ of  island $D_\Delta$. 
    \begin{itemize}
    \item We say that $\beta$ is  \emph{${v}_\gamma$-curve}  if it is a graph of
        a monotonic $\gamma$-Lipschitz function $u = v(u')$ and its type of monotonicity coincides with the functions $u = v_{\pm}(u')$ that correspond to the $\beta^{\pm}$ boundaries of the $\gamma$-island $D_\Delta$. 
    \item We say that  $\beta$ is increasing / decreasing if $\beta$ is a
        graph of increasing / decreasing function $u = f(u')$.
    \end{itemize}
\label{def:v-curve}
\end{dfn}
\begin{dfn}
    We call \emph{$v_{\gamma}$-strip} an open subset of $\gamma$-island $D_\Delta$ bounded by two $v_{\gamma}$-curves.
\label{def:v-strip}
\end{dfn}
\begin{remn}
A $\gamma$-island is both   $h_\gamma$-strip and $v_\gamma$- strip.
\end{remn}
\begin{remn}
    It is important that the type of strip ($h_\gamma$-strip or $v_\gamma$-strip) is defined {\em not} by the the type of monotonicity of its borders (increasing  or decreasing) but the type of boundaries that it connects ($\alpha^{\pm}$ or $\beta^{\pm}$). 
\end{remn}
Figure~\ref{fig:Island(def)} illustrates the definitions of a $\gamma$-island, $h$-curve, $v$-curve, $h$-strip, and $v$-strip.

\begin{dfn} Assume that 
\begin{itemize}
    \item for any $\Delta\in\mathbb{S}^1$ there exists  a $\gamma$-island $D_\Delta\subseteq \mathscr{U}_\Delta$;
    \item  when $\Delta\in \mathbb{S}^1$ varies,  the $\gamma$-island  $D_\Delta$ undergoes a continuous deformation in such a way that its boundaries $\alpha_\Delta^\pm$ and $\beta_\Delta^\pm$ conserve the type of monotonicity.
\end{itemize}
Then we call \emph{ $\gamma$-donut} the set
\begin{gather*}
    T=\bigcup_{\Delta\in\S}D_\Delta.
\end{gather*}
\end{dfn}


%

\begin{figure}
\centering
\includegraphics[width=\linewidth]{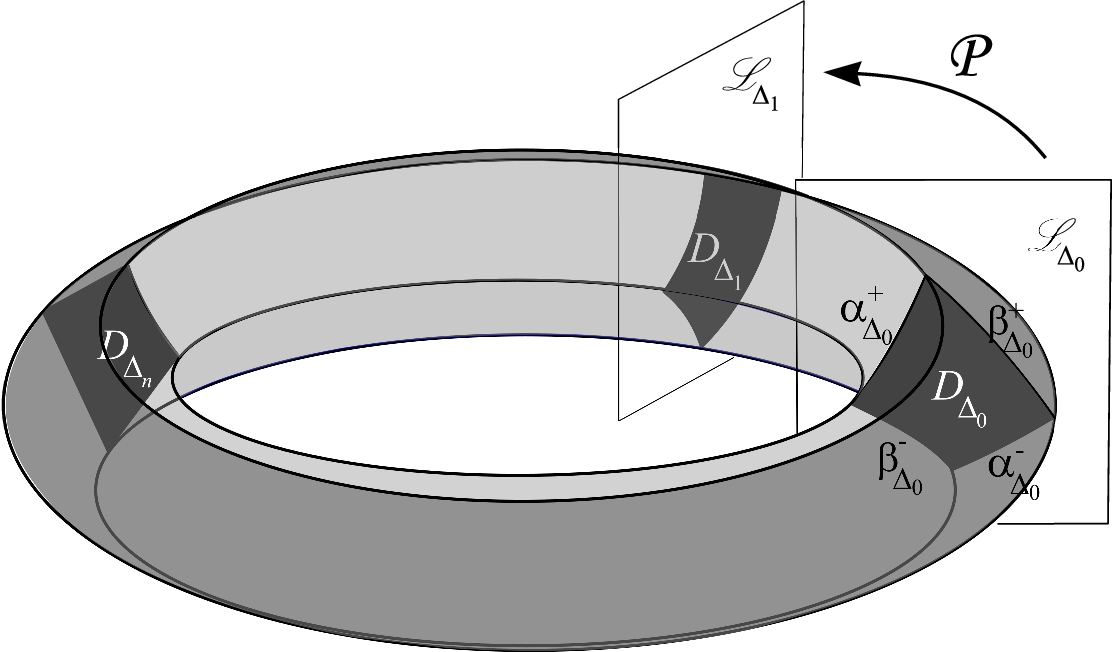}
\caption{A $\gamma$-donut. For each $\Delta$, the intersection of the $\gamma$-donut and the plane $\mathcal{L}_\Delta$ is  a $\gamma$-island. 
}
\label{fig:Donut(def)}
\end{figure}

It follows from the  definition that $T\subseteq \mathscr{U}$. For any ${\bf p}\in T$ both $\mathcal{P}({\bf p})$ and $\mathcal{P}^{-1}({\bf p})$ are defined. The definition of a $\gamma$-donut is illustrated in Fig.~\ref{fig:Donut(def)}.
\begin{remn}
    {If $T$ is a $\gamma_1$-donut and $\gamma_2 > \gamma_1$	then $T$  is also a $\gamma_2$-donut.}
\end{remn}

\begin{dfn}
  An \emph{alphabet} ${\mathcal A}_N$ is a finite or countable set of different indices ${i_k}$, also called \emph{letters}. Here, $N$ denotes the size of the alphabet, which can be a finite integer or (countably) infinite.
\label{def:alphabet}
\end{dfn}
%
%
\begin{dfn}
    Let ${\mathcal A}_N$ be an alphabet {and $\gamma>0$ be   fixed}.
    Define \emph{$\gamma$-donut set} as a set $\mathcal{T} = \bigcup_{i \in {\mathcal A}_N} T^{i}$
    that is a finite or countable union of disjoint $\gamma$-donuts.
\label{def:island-set}
\end{dfn}
\begin{remn}
    Evidently, if $N$ is finite and  $T^{i}$ are $\gamma_i$-donuts, $i\in {\mathcal A}_N$,  then $\mathcal{T} = \bigcup_{i \in {\mathcal A}_N} T^{i}$ is a $\gamma$-donut set for $\gamma=\max_{i\in {\mathcal A}_N}\gamma_i$.     Generically, when $N=\infty$ the set $\mathcal{T}$ may not be a $\gamma$-donut set for any finite $\gamma$ even if each individual $T^{i}$ is a $\gamma_i$-donut.
\end{remn}

\subsection{Width of the Strips}\label{sec:Width}

\begin{figure}
\centering
\includegraphics[width=\linewidth]{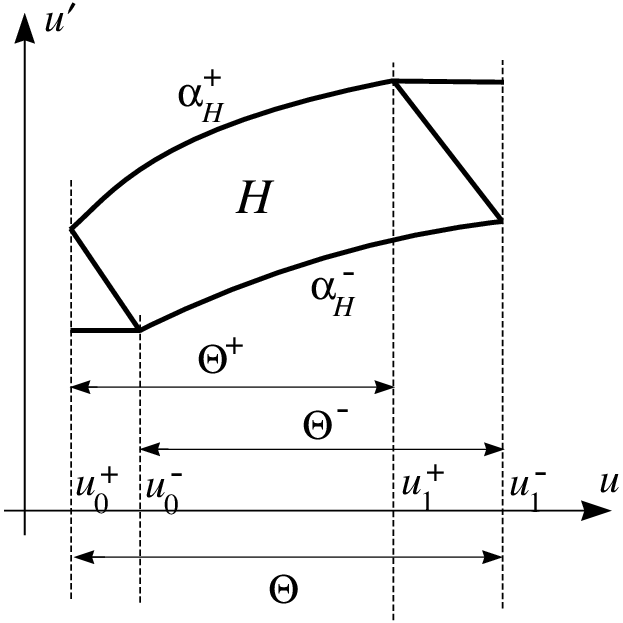}
\caption{On the definition of thickness of a strip}
\label{fig:width}
\end{figure}

Let an $h_\gamma$-strip $H$ lie inside a $\gamma$-island $D_\Delta\subset\mathscr{U}_\Delta$ for some $\Delta$. $H$ is bounded by $h_\gamma$-curves $\alpha^+_H$ and $\alpha^-_H$. These curves are graphs of  functions   $u' = h_{\pm}(u)$. By definition, $h_{\pm}(u)$ are $\gamma$-Lipschitz functions of the same monotonicity type. Let 
$\Theta^{\pm}$ denote the domains of $h_{\pm}(u)$ (see Fig.~\ref{fig:width}). 
Generically, $\Theta^+$ and $\Theta^-$ do not coincide.
Let $\Theta^+ = [u_0^+; u_1^+]$, $\Theta^- = [u_0^-; u_1^-]$ and 
\begin{gather*}
    u_0=\min(u_0^-,u_0^+),\quad u_1=\max(u_1^-,u_1^+).
\end{gather*}
Introduce the new functions $\widetilde{h}_{\pm}(u)$ as follows
\begin{equation}
	\widetilde{h}_{\pm}(u) = \begin{cases}
		h_{\pm}(u_0^{\pm}) & u_0\leq u \leq u_0^{\pm}; \\
		h_{\pm}(u) & u \in \Theta^{\pm}; \\
		h_{\pm}(u_1^{\pm}) & u_1^{\pm}\leq u\leq u_1.
	\end{cases}
\label{eq:continuation}
\end{equation}
Both  functions $\widetilde{h}_{\pm}(u)$ are defined on the same domain $\Theta = \Theta^+ \cup \Theta^-$. Since $h_{\pm}(u)$ are the $\gamma$-Lipschitz functions,  $\widetilde{h}_{\pm}(u)$ are also $\gamma$-Lipschitz on $\Theta$. Let $\widetilde{\alpha}^{\pm}_H$ denote the curves which are  the graphs of $\widetilde{h}_{\pm}(u)$.

\begin{dfn}
\label{def:h-thickness}
	The \emph{thickness} of an ${h}_\gamma$-strip $H$, denoted by
    $d_h(H)$, is the $C$-norm distance between the curves $\widetilde{\alpha}^{\pm}_H$, i.e.
	\begin{equation}
		d_{ h}(H) = \max \limits_{u \in \Theta} |\widetilde{h}_+(u) - \widetilde{h}_-(u)|.
	\label{eq:h-thickness}
	\end{equation}
\end{dfn}

The thickness of $v_\gamma$-strips is defined in the same way.
Let a $v_\gamma$-strip $V$ lie inside a $\gamma$-island $D_\Delta$. Let  $V$  be  bounded by the $v_\gamma$-curves $\beta^+_V$ and $\beta^-_V$.
These curves are the graphs of functions   $u = v_{\pm}(u')$.
Denote domains of these functions by $\Theta^{\pm}$ and 
continue the functions $v_{\pm}(u')$ to the whole interval $\Theta = \Theta^+ \cup \Theta^-$ in the same way as for $h_\gamma$-strips. Introduce  the new functions $\widetilde{v}_{\pm}(u')$ and curves $\widetilde{\beta}^{\pm}_V$.

\begin{dfn}
\label{def:v-thickness}
	The \emph{thickness} of a ${v}_\gamma$-strip $V$, denoted by $d_{\rm v}(V)$, is the $C$-norm distance between the curves $\widetilde{\beta}^{\pm}_V$, i.e.
	\begin{equation}
		d_{ v}(V) =  \max \limits_{u' \in \Theta} |\widetilde{v}_+(u') - \widetilde{v}_-(u')|.
	\label{eq:v-thickness}
	\end{equation}
\end{dfn}


\section{SYMBOLIC DYNAMICS}\label{sec:SymbDyn}

In this section, we formulate sufficient conditions for coding the regular solutions of  Eq.~(\ref{eq:stationary_delta}), i.e., $\ddot{u} =G_\delta(x,u)$, with bi-infinite sequences of letters from a certain alphabet. These conditions are formulated using the auxiliary family of equations given by \eqref{eq:stationary_Delta}, i.e., $\ddot{u} =G_\Delta(x,u)$, where $\Delta \in \S$ is the parameter.
First, we formulate the following two hypotheses.

\begin{hyp}
\label{hypothesis:island-set}
     The set $\mathscr{U}$ includes a $\gamma$-donut set $\mathcal{T}= \bigcup_{i \in \mathcal{A}_N} T^{i}$ for some $\gamma\geq 0$, where $\mathcal{A}_N$ is an alphabet of $N$ symbols. In addition,  there exists a constant $M$ that for any $\Delta \in \S$ and any $i \in \mathcal{A}_N$ all islands  $D_\Delta^i = T^i \cap \mathcal{L}_\Delta$ satisfy the following properties:   $d_{\mathrm{h}}(D^{i}_\Delta) \le M$ and $d_{\mathrm{v}}(D^{i}_\Delta) \le M$.

\end{hyp}
\begin{hyp}
\label{hypothesis:strips-mapping}


    There exists $\gamma\geq 0$ such that:
    \begin{itemize}
    \item For each $i, j \in \mathcal{A}_N$, for any $\Delta\in \S$ and for any  $h_\gamma$-strip $H \subset D^{i}_\Delta$ the set $\widetilde{H}_j \equiv \mathcal{P}(H) \cap D^{j}_{\Delta \oplus 2\theta\pi}$  is an $h_\gamma$-strip, and there exists $\rho_h > 1$, the same for all $i, j \in \mathcal{A}_N$ and for $\Delta\in \S$, such that
        \begin{equation}
	   	d_{{h}}(\widetilde{H}_j) \le  d_{{h}}(H)/\rho_h.
        \end{equation}
    \item For each $i, j \in \mathcal{A}_N$, for any $\Delta\in \S$  and for any $v_\gamma$-strip $V \subset D^{j}_\Delta$ the set $\widetilde{V}_i \equiv \mathcal{P}^{-1}(V) \cap D^{i}_{\Delta \ominus 2\theta \pi}$  is a  ${v}_{\gamma}$-strip, and there exists $\rho_v > 1$, the same for all $i, j \in \mathcal{A}_N$ and for $\Delta\in \S$, such that
        \begin{equation}
		  d_{{v}}(\widetilde{V}_i) \le  d_{{v}}(V)/\rho_v.
        \end{equation}
    \end{itemize}
\end{hyp}
%

\begin{figure*}[ht]
\centering
\includegraphics[width=\linewidth]{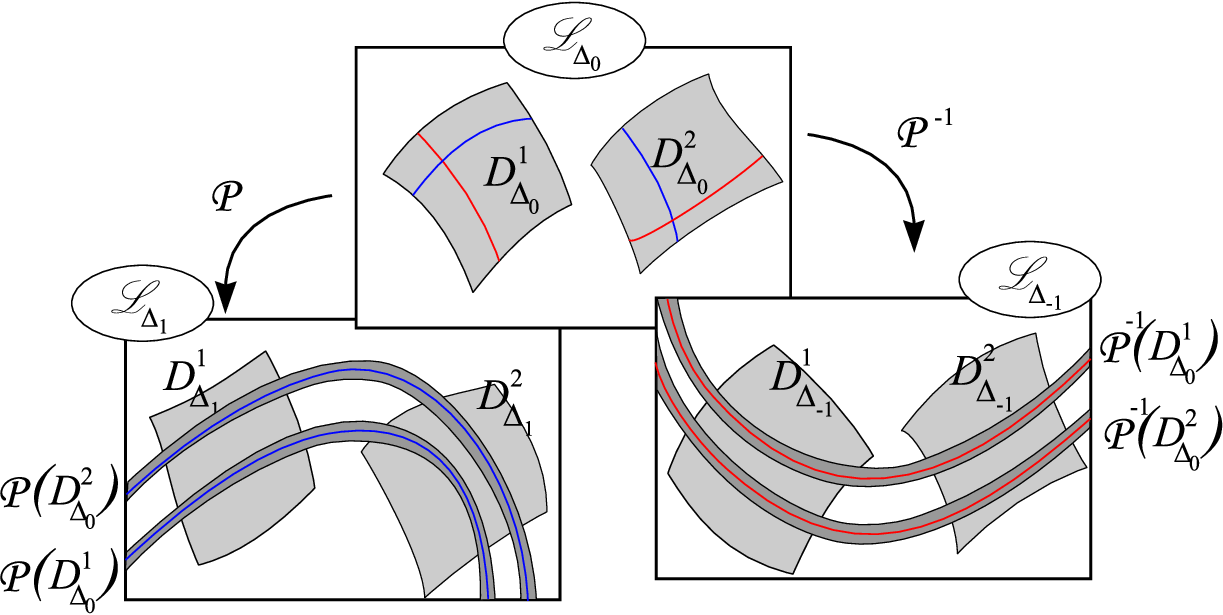}
\caption{A sketch illustrating    Hypothesis~\ref{hypothesis:strips-mapping}. A $\gamma$-donut set consists of two donuts, ${T}^{1}$ and ${T}^{2}$. The plane $\mathcal{L}_{\Delta_0}$ contains  the islands $D_{\Delta_0}^{1}$ and $D_{\Delta_0}^{2}$ from $T^{1}$ and $T^{2}$, respectively.  The planes  $\mathcal{L}_{\Delta_{\pm 1}}$ contain the   corresponding islands at   $\Delta_{1} = \Delta_0\oplus 2\theta \pi$ and $\Delta_{-1} = \Delta_0 \ominus 2\theta \pi$. The images ${\cal P}(D_{\Delta_0}^{1})$ and ${\cal P}(D_{\Delta_0}^{2})$ intersect the $\gamma$-islands $D_{\Delta_{1}}^{1,2}$ and  each intersection is an $h$-strip. Similarly, the pre-images ${\cal P}^{-1}(D_{\Delta_0}^{1})$ and ${\cal P}^{-1}(D_{\Delta_0}^{2})$ intersect the $\gamma$-islands $D_{\Delta_{-1}}^{1,2}$, and  each intersection is a  $v$-strip.}
\label{fig:CompleteSet}
\end{figure*}

\begin{remn}
    In particular, Hypothesis \ref{hypothesis:strips-mapping} implies that both $\mathcal{P}$-image and $\mathcal{P}^{-1}$-images of any $\gamma$-donut from $\mathcal T$ intersect itself and any other $\gamma$-donut from $\mathcal T$. Hypothesis \ref{hypothesis:strips-mapping} is illustrated in Fig.~\ref{fig:CompleteSet}.
\end{remn}

\begin{remn}
    If Hypotheses~\ref{hypothesis:island-set} and \ref{hypothesis:strips-mapping} hold for Eq.~(\ref{eq:stationary_delta}) with some $\delta \in \S$, then these hypothesis hold for Eq.~(\ref{eq:stationary_delta})  with \emph{any} $\delta \in \S$.
\end{remn}

    Let ${\mathcal T}$ be a $\gamma$-donut set. We use    $\mathcal{O}$ to denote the set of all bi-infinite orbits  lying completely in ${\mathcal T}$, such that $\vb{p}_0\in\mathcal{L}_\delta$. Hence
    \begin{gather*}
        \vb{r} \in \mathcal{O}\ \Leftrightarrow\ \vb{r} = \{ \vb{p}_n \}\subset {\mathcal T},\ \mathcal{P}(\vb{p}_n) = \vb{p}_{n+1},\  \vb{p}_0\in\mathcal{L}_\delta,\  n \in \mathbb{Z}.
    \end{gather*}
The set $\mathcal{O}$ has the structure of a metric space, where the distance between $\vb{v}, \vb{w} \in \mathcal{O}$, $\vb{v} = \{ \vb{p}_n \}$,  $\vb{w} = \{ \vb{q}_n \}$ is  the  Euclidean distance between the points $\vb{p}_0$ and $\vb{q}_0$ in $\mathcal{L}_\delta$. 
%

    We use  ${\mathcal R}_N$ to denote  the set of all bi-infinite sequences $\{ \dots, i_{-1}, i_0, i_1, \dots \}$ where $i_k$, $k \in \mathbb{Z}$, are letters of an alphabet $\mathcal{A}_N$ of $N$ symbols  (note that $N$ can be finite or countably infinite, see Definition \ref{def:alphabet}).
The set $\mathcal{R}_N$ has the structure of a topological space, where a system of nested neighborhoods $W_k(\omega^*)$ of an element $\omega^* = \{ \dots, i_{-1}^*, i_0^*, i_1^*, \dots \} \in \mathcal{R}_N$ is defined as 
\begin{gather*}
    W_k(\omega^*) = \{ \omega \ | \ i_s^* = i_s, |s| < k \}, \quad k=1,2,\ldots.
\end{gather*}

The following theorem establishes that the  two hypotheses formulated above provide sufficient conditions for a {\em homeomorphism} between ${\mathcal O}$ and ${\mathcal R}_N$. 
\begin{teo}
	Let Hypotheses  \ref{hypothesis:island-set} and \ref{hypothesis:strips-mapping} be satisfied.
    Define a map $\mathcal{C}: \mathcal{O} \to \mathcal{R}_N$ as follows: $\mathcal{C}(\vb{r}) = \vb{s}$, $\vb{r} \in \mathcal{O}$, $\vb{r} = \{ \vb{p}_k \}$ and $\vb{s} \in \mathcal{R}_N$, $\vb{s} = \{ i_k \}$, such that $i_k$ is the index of the $\gamma$-donut   $T^{i_k} \subset \mathcal{T}$ where the point $\vb{p}_k$ lies. Then $\mathcal{C}$ is a homeomorphism.
\label{thm:coding}
\end{teo}
\begin{remn}
Theorem \ref{thm:coding} generalizes Theorem~1 of Ref.~\cite{AL24} to the quasiperiodic setting. The proof relies on the hyperbolic dynamics of ${\mathcal P}$:  due to Hypothesis \ref{hypothesis:strips-mapping} this map contracts strips in one direction while expanding them in another. Such results are standard in symbolic dynamics, with classical references including Refs.~\cite{M73,GH83,W03}. \ga{Theorem~\ref{thm:coding} holds regardless of whether the parameter $\theta$  in the definition of $G_\delta(x,u)$ is rational or irrational.}
\end{remn}

Theorem  \ref{thm:coding} establishes a correspondence between  regular on $\mathbb{R}$ solutions of Eq.~\eqref{eq:stationary_delta} and  bi-infinite sequences from ${\mathcal R}_N$. If 
${\mathcal T}={\mathscr U}$,  the coding is complete, covering \emph{all} {real-valued} and regular on $\mathbb{R}$ solutions $u(x)$. If ${\mathcal T}\ne {\mathscr U}$, ${\mathcal T}\subset {\mathscr U}$, then the coding may be incomplete, potentially omitting some  regular solutions. 

\begin{proofn}
    Evidently, for each bi-infinite orbit $\vb{r} \in \mathcal{O}$ 
    the image $\vb{s} = \mathcal{C}(\vb{r})$, $\vb{s} \in \mathcal{R}_N$, is defined uniquely.	Let us prove that for each sequence $\vb{s} \in \mathcal{R}_N$ there exists a unique orbit $\vb{r} \in \mathcal{O}$ such that $\vb{s} = \mathcal{C}(\vb{r})$.
	
    Let a sequence $\vb{s} = \{ \dots, i_{-1}, i_0, i_1, \dots \}$ be given. Find locations of the points   $\vb{p} \in D_\delta^{i_0}\subset T^{i_0}$  such that $\mathcal{P}^{-1}(\vb{p}) \in T^{i_{-1}}$, $\mathcal{P}^{-2}(\vb{p}) \in T^{i_{-2}}$, etc. By definition of $\mathcal{P}$, 
    \begin{gather*}
     \mathcal{P}^{-n}({\vb{p}})\in D_{\delta\ominus 2\theta n\pi}^{i_{-n}}\subset T^{i_{-n}},  
    \end{gather*}
     where  $D_{\delta\ominus 2\theta n\pi}^{i_{-n}}$ are $\gamma$-islands, $n=0,1,\ldots$. By Hypothesis~\ref{hypothesis:island-set} each of these $\gamma$-islands satisfies the condition
    \begin{equation}
    d_{\mathrm{h}}\left(D_{\delta\ominus2\theta n\pi}^{i_{-n}}\right) \le M,
    \end{equation}
    and can be regarded as an $h_\gamma$-strip.     Initialize a recursive  procedure as follows.  Since $i_0$ stands on the zero place in $\vb{s}$ then $\vb{p} \in D_\delta^{i_0}$ where $D_\delta^{i_0}$ is a $\gamma$-island. 
    The points $\vb{p} \in T^{i_0}$ such that $\mathcal{P}^{-1}(\vb{p}) \in T^{i_{-1}}$ are situated in the set $H_{i_{-1}, i_0} \equiv \mathcal{P}(D_{\delta\ominus2\theta\pi}^{i_{-1}}) \cap D_\delta^{i_0}$.	Due to Hypothesis~\ref{hypothesis:strips-mapping}, the set $H_{i_{-1}, i_0}$ is an ${h}_{\gamma}$-strip.	Its thickness satisfies the inequality
    \begin{equation}
		d_{{h}}(H_{i_{-1}, i_0}) \le \rho^{-1}_h d_{{h}}(D_{\delta\ominus2\theta\pi}^{i_{-1}}) \le \rho^{-1}_h M.
    \end{equation}
    The points $\vb{p} \in D_\delta^{i_0}$ such that $\mathcal{P}^{-1}(\vb{p}) \in D_{\delta\ominus2\theta\pi}^{i_{-1}}$, $\mathcal{P}^{-2}(\vb{p}) \in D_{\delta\ominus4\theta\pi}^{i_{-2}}$ are situated in the set $H_{i_{-2}, i_{-1}, i_0} \equiv \mathcal{P}(H_{i_{-2}, i_{-1}}) \cap D_\delta^{i_0}$ where $H_{i_{-2}, i_{-1}} = \mathcal{P}(D_{\delta\ominus4\theta\pi}^{i_{-2}}) \cap D_{\delta\ominus2\theta\pi}^{i_{-1}}$.
    By Hypothesis~\ref{hypothesis:strips-mapping}, the  set $H_{i_{-2}, i_{-1}}$ is an $h_\gamma$-strip.
    Hence, the set $H_{i_{-2}, i_{-1}, i_0}$ is an ${h}_{\gamma}$-strip and its thickness satisfies the inequality
    \begin{equation*}
	d_{{h}}(H_{i_{-2}, i_{-1}, i_0}) \le \rho^{-1}_h d_{\mathrm{h}}(H_{i_{-2}, i_{-1}}) \le 
        \rho^{-2}_h d_{{h}}\left(D_{\delta\ominus4\theta\pi}^{i_{-2}}\right) \le \rho^{-2}_h M.
    \end{equation*}
    Continuation of this process yields the sequence of nested ${h}_{\gamma}$-strips,
    \begin{equation}
		D_\delta^{i_0}  \supseteq H_{i_{-1}, i_0} \supseteq H_{i_{-2}, i_{-1}, i_0} \supseteq \dots.
    \label{eq:nested-h-strips}
    \end{equation}
    The thicknesses of these strips tend to zero. The  boundaries of the strips are  the graphs of $\gamma$-Lipschitz functions, so, since the space of the Lipschitz functions is complete 
    (see Remark after Definition \ref{def:Lipschitz}), 
    the intersection
    \begin{gather*}
        \alpha_\infty\equiv\bigcap\limits_{k=0}^\infty H_{i_{-k},\ldots,i_{-1}, i_0}\subset D_\delta^{i_0}
    \end{gather*}
    is an $h_\gamma$-curve.



    Consider the sequence of nested ${v}_{\gamma}$-strips constructed in the same manner,
    \begin{equation}
		D_\delta^{i_0}  \supseteq V_{i_0, i_1} \supseteq V_{i_0, i_1, i_2} \supseteq \dots,
    \label{eq:nested-v-strips}
    \end{equation}
    where $V_{i_{0}, i_1} \equiv \mathcal{P}^{-1}(D_{\delta\oplus2\theta\pi}^{i_{1}}) \cap D_\delta^{i_0}$, $V_{i_0, i_1, i_2} \equiv \mathcal{P}^{-1}(V_{i_1, i_2}) \cap D_{\delta\oplus2\theta \pi}^{i_1}$ where $V_{i_1, i_2}\equiv \mathcal{P}^{-1}(D_{\delta\oplus4\theta\pi}^{i_{2}}) \cap D_{\delta\oplus2\theta\pi}^{i_1}$. Their thicknesses are bounded by the sequence $\{ \rho^{-n}_v M \}$, $n = 0, 1, 2, \dots$, which converges to zero.  By the same reasoning as for $h_\gamma$-strips, the intersection 
    \begin{gather}
        \beta_{\infty}\equiv \bigcap\limits_{k=0}^\infty V_{i_{0},i_1,\ldots,i_{k}}  \subset D_\delta^{i_0}
    \end{gather}
    is a $v_\gamma$-curve.
    
    The orbit $\vb{r} \in \mathcal{O}$ corresponding to the bi-infinite sequence $\{ \dots, i_{-1}, i_0, i_1, \dots \}$ is generated by $\mathcal{P}$- and $\mathcal{P}^{-1}$-iterations of the intersection $\alpha_{\infty} \cap \beta_{\infty}$ which, according to the definitions of $h_\gamma$- and $v_\gamma$-curves, consists of one point. Therefore the orbit $\vb{r}$ exists and is unique.
	
    Let us show that the maps $\mathcal{C}$ and $\mathcal{C}^{-1}$ are continuous. Since $\mathcal{P}$ is continuous, if $\vb{r}^{(1)}, \vb{r}^{(2)} \in \mathcal{O}$,
    \begin{eqnarray*}
		& \vb{r}^{(1)} = \{ \dots, \vb{p}_{-1}^{(1)}, \vb{p}_0^{(1)}, \vb{p}_1^{(1)}, \dots \}, \\
		& \vb{r}^{(2)} = \{ \dots, \vb{p}_{-1}^{(2)}, \vb{p}_0^{(2)}, \vb{p}_1^{(2)}, \dots \},
    \end{eqnarray*}
    are close enough in $\mathcal{O}$ (i.e. points $\vb{p}_0^{(1)}$ and $\vb{p}_0^{(2)}$ are close in the plane $\mathcal L_\delta$), then their $\mathcal{C}$-images share the same central block $|k| < n$ for some $n$.
    Therefore they are also close in $\mathcal{R}_N$-topology.
    Conversely, if $\vb{s}^{(1)} = \mathcal{C}(\vb{p}^{(1)})$ and $\vb{s}^{(2)} = \mathcal{C}(\vb{p}^{(2)})$ share the
    same central block $|k| < n$ for some $n$ then the points $\vb{p}_0^{(1)}$ and $\vb{p}_0^{(2)}$ are
    situated in the same domain $H_{i_{-k}, \dots, i_0} \cap V_{i_0, \dots, i_k}$, so $\vb{p}_0^{(1)}$ and
    $\vb{p}_0^{(2)}$ are close in $\mathcal{O}$-topology. The theorem is proved. $\square$
\end{proofn}


\section{STRIP MAPPING THEOREMS}
\label{sec:StripMap}


Hypotheses  \ref{hypothesis:island-set} and \ref{hypothesis:strips-mapping} enable the  description of regular solutions for Eq.~(\ref{eq:stationary_delta}) in terms of coding sequences. However, verifying these hypotheses is challenging, even numerically. {This is especially true for Hypothesis \ref{hypothesis:strips-mapping} , as it requires computing the $\mathcal{P}$-image of every $h$-strip  and $\mathcal{P}^{-1}$-image of every $v$-strip from the given $\gamma$-donut set. To simplify the numerical verification  of the analogous hypothesis} in the periodic setting, in Ref.~\cite{AL24} it was suggested to analyze of the matrix of linearization for the  Poincar\'e map. Two theorems, referred to as \emph{strip mapping theorems},  reduced the verification of Hypothesis~\ref{hypothesis:strips-mapping}  to  a simple scanning procedure.

The strip mapping theorems can also be generalized to the quasiperiodic setting. 
Let Hypothesis \ref{hypothesis:island-set} hold. Let  the map $\mathcal{P}$ and its inverse $\mathcal{P}^{-1}$ be defined on a $\gamma$-donut set $\mathcal{T} = \bigcup_{i \in {\mathcal A}_N} T^{i}$ where ${\mathcal A}_N$ is an alphabet,  $N<\infty$. For any $\Delta\in\S$ and $i,j\in {\mathcal A}_N$
define 
\begin{align*}
&V_{i,j}(\Delta)={\mathcal P}^{-1}(D_{\Delta\oplus2\theta\pi}^{j})\cap D_{\Delta}^{i},\\[2mm]
&H_{i,j}(\Delta)={\mathcal P}(D_{\Delta\ominus2\theta\pi}^{i})\cap D_{\Delta}^{j}.
\end{align*}

\begin{remn}
The sets $V_{i,j}$ and $H_{i,j}$ are related  as follows
\begin{align}
{\mathcal{P}}(V_{i,j}(\Delta)) = D_{\Delta\oplus2\theta\pi}^{j} \cap {\mathcal{P}}(D^i_\Delta) = H_{i,j}(\Delta\oplus2\theta\pi)\label{V_to_H}\\[2mm]
{\mathcal{P}}^{-1}(H_{i,j}(\Delta)) = D_{\Delta\ominus2\theta\pi}^{i}\cap \mathcal{P}^{-1}( D^j(\Delta)) = V_{i,j}(\Delta\ominus2\theta\pi).\label{H_to_V}
\end{align}
Since Eq.~\eqref{eq:stationary_Delta} does not include first derivatives, for any fixed $\Delta\in \S$  the map ${\mathcal P}$ is  an area-preserving diffeomorphism from  $V_{i,j}(\Delta)\subset {\mathcal L}_\Delta$ onto $H_{i,j}(\Delta\oplus2\theta\pi)\subset{\mathcal L}_{\Delta\oplus2\theta\pi}$, $i,j\in {\mathcal A}_N$.
Denote by ${\mathcal D}{\mathcal P}_{\bf p}$ the linearization of ${\mathcal P}$ at ${\bf p}\in V_{i,j}(\Delta)$, $i,j\in{\mathcal A}_N$. Then for ${\bf p}\in V_{i,j}(\Delta)$ the linear operator ${\mathcal D}{\mathcal P}_{\bf p}$ is represented by $2\times 2$ matrix $(a_{mn})$ and  
\begin{gather}
\det \left[\mathcal{D}\mathcal{P}_{\bf{p}}\right] = 1.
\label{Eq:Area_pres_P}
\end{gather}
Similarly, ${\mathcal P}^{-1}$ is also 
area-preserving, and for any ${\bf q}\in H_{i,j}(\Delta)$, $i,j\in{\mathcal A}_N$,
\begin{gather}
\det \left[\mathcal{D}\mathcal{P}^{-1}_{\bf{q}}\right] = 1,
\label{Eq:Area_pres_Pinv}
\end{gather}
where ${\mathcal D}{\mathcal P}^{-1}_{\bf q}$ is  the linearization of ${\mathcal P}^{-1}$ at ${\bf q}$.
\end{remn}


\begin{teo}[On the mapping of  h-strips]
	\label{thm:h-strips-mapping}
	  Let Hypothesis \ref{hypothesis:island-set} hold and ${\mathcal A}_N$ be an alphabet,   $N<\infty$. Let for any $\Delta\in\S$ and for any pair  $i,j\in {\mathcal A}_N$ the following two conditions be valid:
      \medskip
      
      {\bf (H-1)}: one of the following requirements holds:
	%
	\begin{enumerate}
		\item[(A)] the borders $\alpha_{i}^{\pm}$ of the island $D_{\Delta}^{i} $ are increasing $\gamma$-curves,    $\forall \vb{p} \in \overline{V_{i,j}(\Delta)}$ the signs of entries $\{ a_{mn} \}$ in the matrix of the linearization ${\mathcal D} \mathcal{P}_{\vb{p}} = (a_{mn})$ have exactly one of the following configurations:
		      \begin{center}
			      (a) $\begin{psm} + & + \\ + & + \end{psm}$, \quad
			      (b) $\begin{psm} - & - \\ - & - \end{psm}$, \quad
			      (c) $\begin{psm} + & + \\ - & - \end{psm}$, \quad
			      (d) $\begin{psm} - & - \\ + & + \end{psm}$
		      \end{center}
		      and at the same time the borders $\alpha_{j}^\pm$ of $D_{\Delta\oplus2\theta\pi}^{j}$ are increasing $\gamma$-curves for cases (a), (b), and decreasing $\gamma$-curves for (c), (d);
		\item[(B)] the borders $\alpha_{i}^{\pm}$ of the island $D_{\Delta}^{i} $ are decreasing $\gamma$-curves, $\forall \vb{p} \in \overline{V_{i,j}(\Delta)}$ signs of $\{ a_{mn} \}$ in the matrix of the lineaization ${\mathcal D} \mathcal{P}_{\vb{p}}$ have exactly one of the following configurations:
		      \begin{center}
			      (a) $\begin{psm} + & - \\ - & + \end{psm}$, \quad
			      (b) $\begin{psm} - & + \\ + & - \end{psm}$,	\quad
			      (c) $\begin{psm} + & - \\ + & - \end{psm}$, \quad
			      (d) $\begin{psm} - & + \\ - & + \end{psm}$
		      \end{center}
		      and at the same time borders $\alpha_{j}^{\pm}$ of $D_{\Delta\oplus2\theta\pi}^{j}$ are decreasing $\gamma$-curves for cases (a), (b), and increasing for (c), (d).
	\end{enumerate}
    \medskip

    {\bf (H-2)}: there exists $\rho_h > 1$ such that for any ${\bf p} \in \overline{V_{i,j}(\Delta)}$, $|a_{11}| \ge \rho_h$.
    \medskip
    
	Then
	\begin{itemize}
		\item[{\bf(H-i)}] for any ${h}_\gamma$-strip $H \subset D_{\Delta}^{i}$ there exists $\gamma_{ij}$
		      (independent on $\Delta$)
		      such that $\mathcal{P} (H) \cap D_{\Delta\oplus2\theta\pi}^{j} = \widetilde{H}^{j}$ is also an
              ${h}_{\gamma_{ij}}$-strip;
		\item[{\bf (H-ii)}] $d_{{h}}(\widetilde{H}^{j}) \le d_{{h}}(H)/\rho_h$.
	\end{itemize}
\end{teo}

\begin{remn}
    The signs ``$+$'' and ``$-$''  mean  that  the \emph{strict} inequalities $a_{mn} > 0$ and $a_{mn} < 0$   must hold, respectively.
\end{remn}

\begin{proofn}
    The proof follows step-by-step the proof of Theorem~2 for periodic case in Ref.~\cite[section~5]{AL24}. The only difference is that in the periodic case, the Poincar\'e map  sends  $\mathbb{R}^2$ to itself, whereas  in our case, its counterpart  $\mathcal{P}$ maps a $\Delta$-slice to a $(\Delta\oplus2\theta \pi)$-slice for each $\Delta \in \S$. Let us   sketch the    proof. At the first step, one shows that under the conditions of the theorem $\mathcal{P}$ preserves Lipschitz and monotonicity properties when mapping the curves from $V_{i,j}(\Delta)\subset D_{\Delta}^{i}$ to $D_{\Delta\oplus2\theta\pi}^{j}$.  At the second step one uses this fact to prove that for any $h_\gamma$-strip $H\subset D^i(\Delta)$ its $\mathcal{P}$ image $\tilde{H}^j\subset D_{\Delta\oplus2\theta\pi}^{j}$ is also  a $h_{\gamma_{ij}}$-strip for some $\gamma_{ij}$. Finally, one shows that the condition (H-2) implies that  $\mathcal{P}$ contracts the $h$-strips. 
    $\square$
\end{proofn}

\begin{teo}[On the mapping of v-strips]
	\label{thm:v-strips-mapping}
	  Let Hypothesis \ref{hypothesis:island-set} hold and ${\mathcal A}_N$ be an alphabet,   $N<\infty$. Let for any $\Delta\in\S$ and for any pair  $i,j\in {\mathcal A}_N$ the following two conditions be valid:
      \medskip
      
      {\bf (V-1)}: one of the following conditions hold:
	%
	\begin{enumerate}
		\item[(A)] the borders $\beta_{j}^{\pm}$ of the island $D_{\Delta}^{j}$ are increasing $\gamma$-curves, $\forall \vb{q} \in \overline{H_{i,j}(\Delta)}$ the signs of entries $\{ b_{mn} \}$ in the matrix of the linear operator ${\mathcal D} \mathcal{P}_{\vb{q}}^{-1} = (b_{mn})$ have exactly one of the following configurations:
		      \begin{center}
			      (a) $\begin{psm} + & + \\ + & + \end{psm}$, \quad
			      (b) $\begin{psm} - & - \\ - & - \end{psm}$, \quad
			      (c) $\begin{psm} + & + \\ - & - \end{psm}$, \quad
			      (d) $\begin{psm} - & - \\ + & + \end{psm}$
		      \end{center}
		      and at the same time the borders $\beta_{i}^{\pm}$ of $D_{\Delta\ominus2\theta\pi}^{i}$ are increasing $\gamma$-curves for cases (a), (b), and decreasing $\gamma$-curves for (c), (d);
		\item[(B)] the borders $\beta_{j}^{\pm}$ of an island $D_{\Delta}^{j}$ are decreasing $\gamma$-curves, $\forall \vb{q} \in \overline{H_{i,j}(\Delta)}$ signs of $\{ b_{mn} \}$ in the matrix of the linear operator ${\mathcal D} \mathcal{P}_{\vb{q}}^{-1} = (b_{mn})$ have exactly one of the following configurations:
		      \begin{center}
			      (a) $\begin{psm} + & - \\ - & + \end{psm}$, \quad
			      (b) $\begin{psm} - & + \\ + & - \end{psm}$,	\quad
			      (c) $\begin{psm} + & - \\ + & - \end{psm}$, \quad
			      (d) $\begin{psm} - & + \\ - & + \end{psm}$
		      \end{center}
		      and at the same time borders $\beta_{i}^{\pm}$ of $D_{\Delta\ominus2\theta\pi}^{i}$ are decreasing $\gamma$-curves for cases (a), (b), and increasing for (c), (d).
	\end{enumerate}
    \medskip
  
    {\bf (V-2)}: there exists $\rho_v > 1$ such that for any ${\bf q} \in \overline{H_{i,j}(\Delta)}$, $|b_{22}| \ge \rho_v$.
    \medskip
    
	Then
	\begin{itemize}
        \item[{\bf (V-i)}] for any ${v}_\gamma$-strip $V \subset D_{\Delta}^{j}$ there exists $\gamma_{ij}$
		      (independent on $\Delta$)
            such that $\mathcal{P}^{-1} (V) \cap D_{\Delta\ominus2\theta\pi}^{i} = \widetilde{V}^{i}$ is also an      ${v}_{\gamma_{ij}}$-strip;
		\item[{\bf (V-ii)}] $d_{\mathrm{v}}(\widetilde{V}^{i}) \le d_{{v}}(V)/\rho_v$.
	\end{itemize}
\end{teo}

The proof of Theorem \ref{thm:v-strips-mapping} repeats the steps of the proof of Theorem \ref{thm:h-strips-mapping} to $v_\gamma$-strips.
\medskip

Theorems \ref{thm:h-strips-mapping} and \ref{thm:v-strips-mapping} are equivalent in the following sense.

\begin{teo}
	\label{thm:v-and_h-relation}
	Let Hypothesis~\ref{hypothesis:island-set} hold and $\mathcal{A}_N$ be an alphabet,  $N<\infty$.	Then the conditions (H-1) and (H-2) of Theorem~\ref{thm:h-strips-mapping} imply the conclusions (V-i) and (V-ii) of Theorem~\ref{thm:v-strips-mapping}. Conversely, the conditions (V-1) and (V-2) of Theorem~\ref{thm:v-strips-mapping} imply the conclusions (H-i) and (H-ii) of Theorem~\ref{thm:h-strips-mapping}.
\end{teo}

\begin{proofn}
  Due to relations (\ref{V_to_H}) and (\ref{H_to_V}), if ${\bf p}\in \overline{V_{i,j}(\Delta)}$, $\Delta\in\S$, then ${\bf q}={\mathcal{P}}({\bf p}) \in \overline{H_{i,j}(\Delta_1)}$ where $\Delta_1=\Delta\oplus2\theta\pi$. Let
   ${\mathcal D}{\mathcal{P}}_{\bf p} =\begin{psm} a_{11} & a_{12} \\ a_{21} & a_{22} \end{psm}$ satisfy the conditions (A) or (B) of Theorem \ref{thm:h-strips-mapping}. Since ${\rm det}~[{\mathcal D}{\mathcal{P}}_{\bf p}]=1$, for ${\bf q}\in \overline{H_{i,j}(\Delta_1)}$
\begin{gather*}
{\mathcal D} {\mathcal{P}}^{-1}_{\bf q} = \left[{\mathcal D}{\mathcal{P}}_{\bf p}\right]^{-1}=\begin{psm} a_{22} & -a_{12} \\ -a_{21} & a_{11} \end{psm}=\begin{psm} b_{11} & b_{12} \\ b_{21} & b_{22} \end{psm}.
\end{gather*}
So, the signs of $(b_{mn})$ agree with conditions for ${\bf q}\in \overline{H_{i,j}(\Delta)}$ of Theorem \ref{thm:v-strips-mapping} for any $\Delta\in\S$. Then the corresponding set of signs for ${\mathcal D} {\mathcal{P}}^{-1}$, either (A) or (B), in Theorem \ref{thm:v-strips-mapping} holds and $|b_{22}|=|a_{11}|\geq\rho_h>1$. The converse statement, that the conditions of Theorem \ref{thm:v-strips-mapping} imply the conclusions of Theorem \ref{thm:h-strips-mapping}, is proved similarly. $\square$
\end{proofn}

By Theorem~\ref{thm:v-and_h-relation}, Hypothesis~\ref{hypothesis:strips-mapping} is satisfied if the conditions of either Theorem~\ref{thm:h-strips-mapping} or Theorem~\ref{thm:v-strips-mapping} hold.
From a practical standpoint, verifying the conditions of either of these theorems is significantly simpler than directly checking Hypothesis~\ref{hypothesis:strips-mapping}. Rather than studying the images of all $h$- and $v$-strips  the analysis reduces to examining the linearization matrices ${\mathcal D}{\mathcal P}_{\bf p}$ or  ${\mathcal D}{\mathcal P}^{-1}_{\bf q}$ at points ${\bf p}, {\bf q}$ within the donut set.

\section{NUMERICAL VERIFICATION OF the HYPOTHESES}\label{sec:NumVer}

The conditions of Hypotheses \ref{hypothesis:island-set} and \ref{hypothesis:strips-mapping} can be checked using  numerical computations. To this end,  we select the scanning domain $\Omega\subset{\mathcal L}$. 
\begin{gather*}
\Omega=\{(\Delta,u,u')|\,0\leq \Delta<2\pi,\,-U\leq u\leq U,\, -U'\leq u'\leq U' \}.
\end{gather*}
The domain $\Omega$ is chosen to be large enough to fully encompass the set ${\mathscr U}$, whenever possible.
We introduce in $\Omega$ a uniform 3D-grid $(\Delta_k,u_l,u'_m)$, $k=1,\ldots, N_\Delta$, $l=-N_u,\ldots N_u$, $m=-N_{u'},\ldots N_{u'}$. 
The computation consists of two steps.

In the first step, we verify the conditions of Hypothesis~\ref{hypothesis:island-set}. We compute a grid approximation  of the set  ${\mathscr U}\cap\Omega$ to obtain  numerical evidence that   ${\mathscr U}$ contains a $\gamma$-donut set. For this purpose, we select points in $\Omega$ for which both ${\mathcal P}$ and ${\mathcal P}^{-1}$  exist.  For each fixed $k$, we take $\Delta=\Delta_k$ and use  $(u_l,u'_m)$, $l=-N_u,\ldots N_u$, $m=-N_{u'},\ldots N_{u'}$ as initial data for the Cauchy problem
\begin{gather}
    \ddot{u} = G_{\Delta_k}(x,u),\quad u(0)=u_l,\quad \dot{u}(0)=u'_m.\label{Eq:Check} 
\end{gather}
The problem (\ref{Eq:Check}) is   solved numerically.
If the solution to (\ref{Eq:Check}) exists on the interval $[0, 2\pi]$, then the image ${\mathcal P}(\mathbf{p})$ is defined for the grid node $\mathbf{p} = (\Delta_k, u_l, u'_m)$.  Thus,  this node belongs to the grid approximation of the set $\mathscr{U}^+_{\Delta_k}\cap\Omega$. Similarly, we obtain the grid approximation  of the set $\mathscr{U}^-_{\Delta_k}\cap\Omega$ by selecting  the nodes for which  the solution to (\ref{Eq:Check}) can be extended to the interval  $[-2\pi;0]$. Then we compute the intersection of these sets, that is $\mathscr{U}_{\Delta_k}\cap\Omega$. Hypothesis  \ref{hypothesis:island-set} holds if for any $\Delta_k$, $k=1,\ldots,N_k$, the set  $\mathscr{U}_{\Delta_k}\cap\Omega$ includes a finite number of $\gamma$-islands (for some $\gamma$  independent of $k$), and the monotonic properties of the islands' boundaries are the same for all $\Delta_k$. 

If the first step is successful, we proceed to the second step: the numerical verification of Hypothesis~\ref{hypothesis:strips-mapping}. The base for this is provided by Theorems  \ref{thm:h-strips-mapping} and  \ref{thm:v-strips-mapping} which apply to the case when the $\gamma$-donut set consists of a finite number of $\gamma$-donuts. Due to Theorem \ref{thm:v-and_h-relation}, it is enough to check the conditions of only one of these theorems. We have used Theorem  \ref{thm:h-strips-mapping}.  On an adapted grid from the prior scan, for any fixed $\Delta_k$ we  solve  the Cauchy problem (\ref{Eq:Check})  and select the grid nodes for which the solution can be extended to the interval $[-2\pi;4\pi]$. For any such node ${\bf p}=(\Delta_k,u_l,u'_m)$ both   ${\mathcal P}^2{\bf p}$ and ${\mathcal P}^{-1}{\bf p}$ exist. Thus this node belongs to the numerical approximation of the set    $\bigcup_{i,j}V_{i,j}(\Delta_k)\cap\Omega$. 
For any node ${\vb{p}}$ of this set we compute the matrix of linearization ${\mathcal D} \mathcal{P}_{\vb{p}}=\begin{psm} a_{11} & a_{12} \\ a_{21} & a_{22} \end{psm}$. We store the signs of its entries together with the value of $a_{11}$. 
Then one has to ensure that the conditions (H-1) and (H-2) hold for any pair $i,j\in{\mathcal A}_N$. 

The procedure is repeated for any $\Delta_k$, $k=1,\ldots, N_\Delta$. If the points (H-1) and (H-2) hold for any $\Delta_k$ we conclude that the conditions of Hypotheses ~\ref{hypothesis:strips-mapping} are valid.

Algorithm~\ref{alg} summarizes the procedure for verification of Hypotheses~\ref{hypothesis:island-set}  and \ref{hypothesis:strips-mapping}.   Illustrations of this procedure are provided in Sec.~ \ref{sec:Examples}.

\begin{algorithm}[H]
\caption{Numerical Check of Hypotheses~\ref{hypothesis:island-set}  and \ref{hypothesis:strips-mapping}}
\label{alg}
\begin{algorithmic}
{
 \STATE \textbf{Input:} The scanning area
\begin{gather*}
\Omega=\{(\Delta,u,u')|\,0\leq \Delta<2\pi,\,-U\leq u\leq U,\, -U'\leq u'\leq U' \}.
\end{gather*}
covered by a uniform 3D grid
\begin{gather*}
    (\Delta_k,u_l,u'_m),~k=1,\ldots, N_\Delta,~ l=-N_u,\ldots N_u,~ m=-N_{u'},\ldots N_{u'}.
\end{gather*}
\\[2mm]
	\STATE \textbf{Step 1: Verify Hypothesis \ref{hypothesis:island-set}}.\\
    \begin{enumerate}
		\setlength{\itemsep}{1pt}
		\setlength{\parskip}{0pt}
  		\setlength{\parsep}{0pt}
		\item[\textbf{(i)}] {\bf Grid approximation of $\mathscr{U}\cap\Omega$}.  For each  $(k,l,m)$:\\
        Solve  Eq.~(\ref{Eq:Check}) with  initial data ${u}(0)=u_l$ $\dot{u}(0)=u'_m$.
        Store the  sets ${\Omega}_k$, $k=1,\ldots, N_\Delta$,  that consist of pairs $(u_l,u'_m)$  corresponding to  solutions $u(x)$ that are regular on $x\in[-2\pi;2\pi]$.
        \item[\textbf{(ii)}] {\bf Searching for islands in $\Delta$-slices}. For each $\Delta_k$, $k=1,\ldots, N_\Delta$, separate connected components in each ${\Omega}_k$, selecting those that correspond to $\gamma$-islands.
        \item[\textbf{(iii)}] {\bf Searching for a donut set}.        
        Select islands that  
        form donuts as $k$ runs from 1 to $N_\Delta$. Introduce $N$ ---  the number of donuts in the donut set, and enumerate the donuts using indices $i$ from an $N$-symbol alphabet $\mathcal{A}_N$. 
    \end{enumerate}
   
	\STATE \textbf{Step 2: Verify Hypothesis \ref{hypothesis:strips-mapping}}.\\

    For each $k=1,\ldots, N_\Delta$ repeat the following steps:
    
    \begin{enumerate}
		\setlength{\itemsep}{1pt}
		\setlength{\parskip}{0pt}
  		\setlength{\parsep}{0pt}
		 
        \item[\textbf{(i)}] For each $i$ from $\mathcal{A}_N$,   and  for each $\textbf{p} = (\Delta_k, u_l, u_m')$ from $D_{\Delta_k}^i$:\\  Solve Eq.~(\ref{Eq:Check}) for initial data ${u}(0)={u}_l$ $\dot{u}(0)={u}_m'$. If the solution $u(x)$ is regular on $[-2\pi, 4\pi]$, then store the following data:
        \begin{itemize}
        \item  the pair $(i,j)$, where $j\in \mathcal{A}_N$ is the index of the donut containing the point $(\Delta_k \oplus 2\pi \theta, u(2\pi), u'(2\pi))$; 
        
        \item the $2\times 2$ linerization matrix  ${\mathcal D} \mathcal{P}_{\vb{p}}=(a_{m,n})_{m,n=1,2}$; 
        \end{itemize}
        
        \item[\textbf{(ii)}] 
        Check that the pairs stored  at the previous step contain every element of $\mathcal{A}_N \times \mathcal{A}_N$;
        
        \item[\textbf{(iii)}] For all stored linearization matrices,   check that the signs of their entries $a_{mn}$  meet  the conditions of Theorem~\ref{thm:h-strips-mapping};
        
         \item[\textbf{(iv)}]  Compute  $\rho_* = \min  \left| a_{11}\right|$ over all stored linearization  matrices and check that $\rho_* > 1$.	
    \end{enumerate}
    If conditions corresponding to steps (ii), (iii), (iv) hold for every $k$, then  Hypothesis~\ref{hypothesis:strips-mapping} is satisfied 
 }
\end{algorithmic}
\end{algorithm}

\section{EXAMPLES} 
\label{sec:Examples}

We exemplify this procedure using Eq. (\ref{StatGPEq}) with the quasiperiodic potential, see the setting (\ref{Eq:CaseA}).
In this case the equation reads  
\begin{equation}
    \label{eq:general-form-equation}
    \ddot{u} + \left(\mu + A_1 \cos{2 x} + A_2 \cos{(2 \theta x + \delta)}\right) u - u^3 = 0. \
\end{equation}
Here the phase shift $\delta$  is fixed, $\delta\in [0, 2 \pi)$,  $\theta>1$ is an irrational number. In the examples below, we use  the golden ratio   $\theta=(\sqrt{5}+1)/2$. In the context of BEC, $\mu$ represents the chemical potential and $A_{1,2}$ are the   amplitudes of the   laser beams that form the quasiperiodic lattice.  The independent variable $x$ has been scaled to make the linear part of Eq.~(\ref{eq:general-form-equation})  consistent with the Mathieu equation. Then the map $\mathcal{P}$ is defined over the period of one of cosines, i.e., for $x\in[0;\pi)$, instead of $x\in[0;2\pi)$. So,  $\mathcal{P}$ is defined as follows:  $\mathcal{P}({\bf p}_0)={\bf p}_1$ where ${\bf p}_0=(\Delta_0,u_0,u_0')$, ${\bf p}_1=({\Delta_1}, u_1, u_1')$ and
     \begin{itemize}
         \item {$\Delta_1 =  \Delta_0 \oplus 2\theta\pi$} ;
         \item $u_1=u(\pi)$, $u_1'=\dot{u}(\pi)$ where $u(x)$ is the solution of the Cauchy problem for Eq.~ (\ref{eq:stationary_Delta}) with $\Delta=\Delta_0$ and initial data $u(0)=u_0$, $\dot{u}(0)=u_0'$. 
    \end{itemize}


   \begin{figure*}
		\centering
        \includegraphics[width=1.0\linewidth]{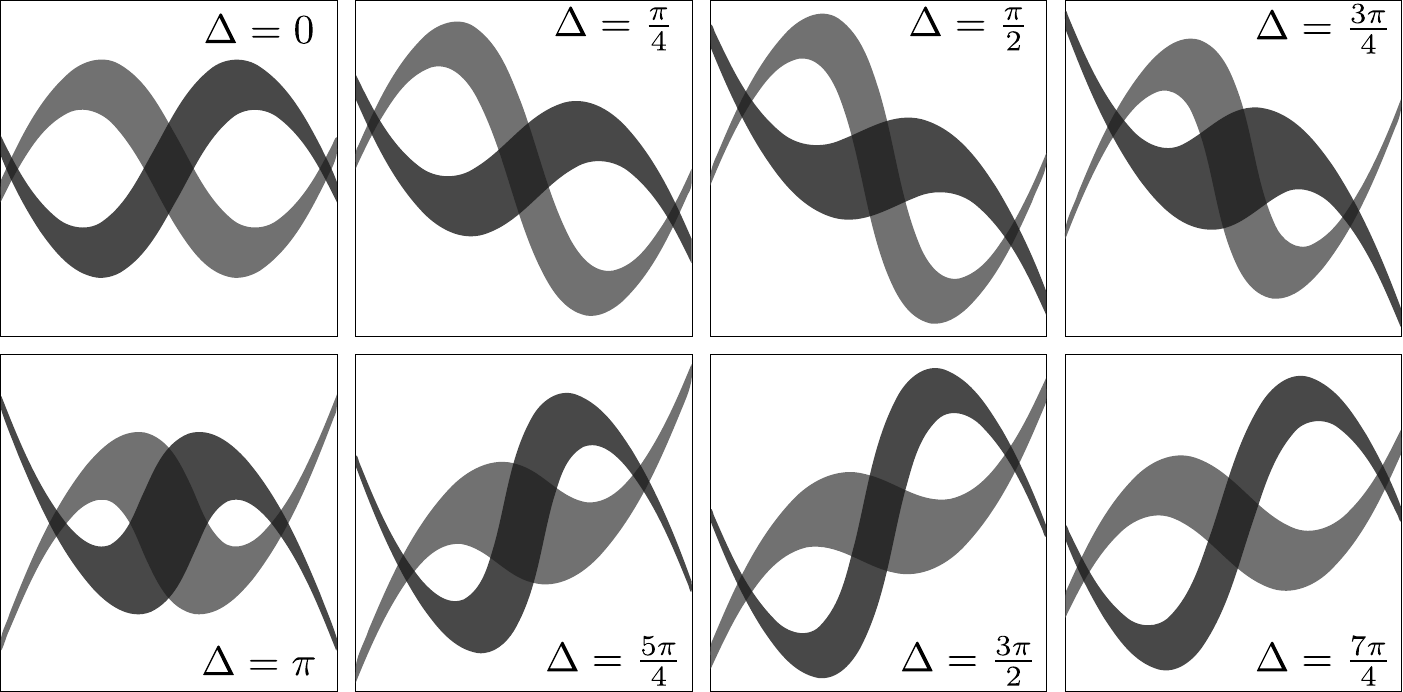}
		\caption{Numerical verification of Hypothesis~\ref{hypothesis:island-set} for Example~\ref{exa:example-1}. Each panel shows the sets ${\mathscr U}^+_{\pi k/4}$ (dark gray), ${\mathscr U}^-_{\pi k/4}$ (light gray) and ${\mathscr U}_{\pi k/4}$ (black). For each $\Delta$, there are  exactly three $\gamma$-islands  and each $\gamma$-island continuously changes with $\Delta$. Each panel shows the region $(u,u')\in [-2;2]\times[-2;2]$.
        }
	    \label{fig:example-1-diagram-grid}
	\end{figure*}

 \begin{figure*}[ht]
		\centering
        \includegraphics[width=1.0\linewidth]{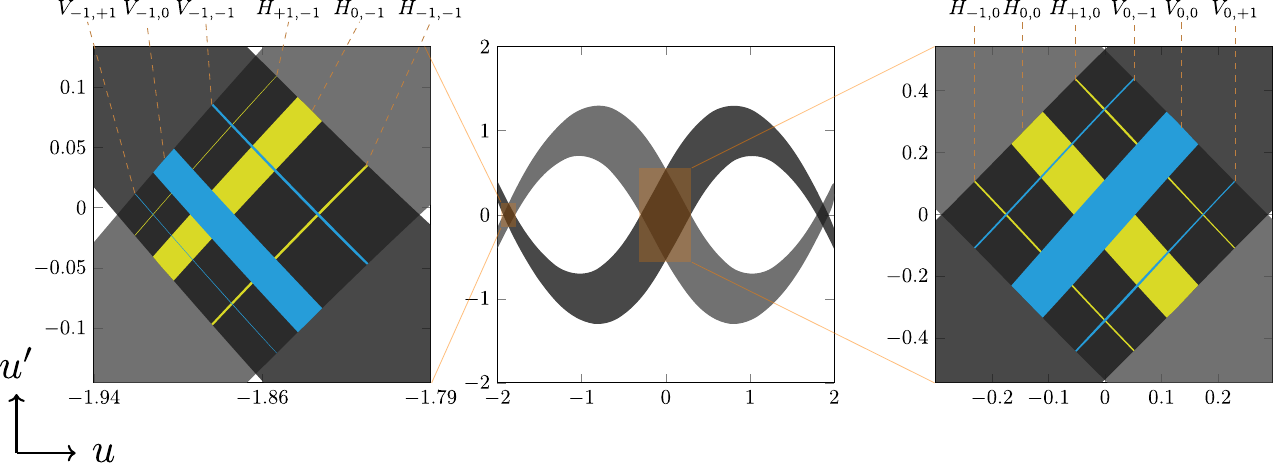}
		\caption{Example~\ref{exa:example-1}: the positions of the islands in $\Delta$-slice with $\Delta=0$ (central panel) and the $v$-strips $ V_{i,j}(0)$ (blue) and $h$-strips $H_{i,j}(0)$ (yellow). Left and right panels show   the island $D_{0}^{-1}$ and   $D_{0}^{0}$, respectively.  The island $D_{0}^{1}$ is not shown, as  it is symmetric to $D_{0}^{-1}$ with respect to the origin.  
        }
	    \label{fig:example-1-diagram-magnifier}
	\end{figure*}

\begin{exa}
    \label{exa:example-1}  
    Consider Eq.~\eqref{eq:general-form-equation} with $\mu = 1$, $A_1 = 3$, $A_2 = 2$ and $\delta=0$. The set ${\mathscr U}$ has been computed by scanning the domain $\Omega\subset{\mathcal L}$ on a grid of $16\times 2000\times2000$ nodes. Figure~\ref{fig:example-1-diagram-grid} shows the sets ${\mathscr U}^+_{\pi k/4}$ (dark gray), ${\mathscr U}^-_{\pi k/4}$ (light gray) and ${\mathscr U}_{\pi k/4}$ (black), $k=0,\ldots,7$. The sets ${\mathscr U}_{\pi k/4}$ are the cross-sections of the set ${\mathscr U}\subset {\mathcal L}$  by the $\Delta$-slices, where $\Delta={\pi k/4}$. Each $\Delta$-slice  contains three $\gamma$-islands ($\gamma$ is finite). We  introduce an  alphabet $\mathscr{A}_3  = \{-1, 0, 1\}$ and  label the  islands as $D_\Delta^{-1}$, $D_\Delta^0$ and $D_\Delta^{1}$. Without loss of generality, we use $D_\Delta^0$ to denote the $\gamma$-island that contains the origin $(u, u') = (0, 0)$. Since for $\delta=0$ the quasiperiodic potential is an even function of $x$, the other two islands, $D_\Delta^{-1}$ and $D_\Delta^{1}$, are mirrored copies of one  another with respect to the origin $(u, u') = (0, 0)$.
    Each of the  islands undergoes a  continuous deformation when $\Delta$ varies, conserving the monotonic properties of its $\alpha$- and $\beta$-boundaries. So, in  $\mathcal L$ each of these three islands generates a $\gamma$-donut. Therefore, we conclude that the set ${\mathscr U}$ contains a $\gamma$-donut set $\mathcal T$.  This yields   numerical evidence supporting Hypothesis \ref{hypothesis:island-set}. Enlarging the scanning area $\Omega$ does not reveal other intersections of ${\mathscr U}^+_{\Delta}$ and ${\mathscr U}^-_{\Delta}$. This indicates that    ${\mathscr U}=\mathcal T$.

    The next step involves the  analysis of $h$- and $v$-strips. To find $V_{i,j}(\Delta)$, $i,j\in\mathscr{A}_3 $, we scan the islands in a fixed $\Delta$-slice identifying points that have  a ${\mathcal P}^2$-image. By definition,  $V_{i,j}(\Delta)$ is a $v$-strip which is situated in the island $D_\Delta^i$ and   mapped by ${\mathcal P}$ to the island $D_{\Delta\oplus 2\pi\theta}^j$.  We also compute $h$-strips $H_{i,j}(\Delta)$ selecting the points in the islands in $\Delta$-slice that have a $ \mathcal{P}^{-2} $-image. By definition,  $H_{i,j}(\Delta)$ is an $h$-strip  which is situated in the island $D_\Delta^j$  and  mapped by ${\mathcal P}^{-1}$ into $D_{\Delta\ominus 2\pi\theta}^i$. Figure~\ref{fig:example-1-diagram-magnifier} shows   the strips for $\Delta=0$, i.,e., $V_{i,j}(0)$ and $H_{i,j}(0)$.

    To check the conditions of Theorem \ref{thm:h-strips-mapping}, for any grid point  ${\bf p}\in V_{i,j}(\Delta)$ and for any $\Delta = \Delta_k\in[0;2\pi)$  we compute   the matrices ${\mathcal D}{\mathcal P}_\mathbf{p}$.
    Since this computation involves only  the points  within the $\gamma$-islands,  we perform this procedure on a finer numerical grid than that used for  verifying Hypothesis~\ref{hypothesis:island-set}.
    The result is shown in Fig.~\ref{fig:example-1-diagram-islands-grids}, where we use different colors to represent different combinations of signs for the entries of the linearization matrices. For each considered $v$-strip $V_{i,j}(\Delta)$, the linearization  satisfies the requirements of Theorem~\ref{thm:h-strips-mapping}.    
    To complete the verification  of the conditions of this theorem, for each island $D_\Delta^{-1}$, $D_\Delta^0$, and $D_\Delta^{1}$, we additionally 
     compute $\rho_h$ as
    \begin{gather*}
       \min_{{\bf p}\in V_{-1,j}(\Delta)}|a_{11}|,\quad \min_{{\bf p}\in V_{1,j}(\Delta)}|a_{11}|,\quad \min_{{\bf p}\in V_{0,j}(\Delta)}|a_{11}|.
    \end{gather*}
    Here $j=-1,0,1$ and $\Delta\in [0;2\pi)$. The result is plotted in Fig.~\ref{fig:example-1-island-1-mu-graph}. For each island, we find $\rho_h > 1$, providing numerical support for the validity of Theorem~\ref{thm:h-strips-mapping} and, consequently, of Hypothesis~\ref{hypothesis:strips-mapping}.
    

    Since both Hypotheses~\ref{hypothesis:island-set} and \ref{hypothesis:strips-mapping} have been numerically verified,  we conclude that there exists a one-to-one correspondence between all bi-infinite sequences over the  three-symbol alphabet $\mathcal{A}_3=\{-1,0,1\}$ and the regular and real-valued solutions of Eq.~\eqref{eq:general-form-equation} for  the given parameters. This correspondence implies that for any sequence $(\ldots,i_{-1},i_0,i_1,\ldots)$, $i_k\in\mathcal{A}_3$, there exists a unique solution of Eq.~\eqref{eq:general-form-equation} such that $(\delta\oplus 2k\pi\theta,u( k\pi),\dot{u}(k\pi))\in T^{i_k}$, $k\in\mathbb{Z}$.
    Moreover, since our computations confirm that ${\mathscr U}=\mathcal T$ we conclude that {\emph all} regular solutions of Eq.~\eqref{eq:general-form-equation} are in one-to-one correspondence with these bi-infinite codes. This means that for any regular solution of Eq.~\eqref{eq:general-form-equation} there exists the coding sequence $(\ldots,i_{-1},i_0,i_1,\ldots)$, $i_k\in\mathcal{A}_3$.
    
    Figure~\ref{fig:solitons}(a) shows four solutions that correspond to the codes 
    $(\ldots,\underline{0},1,0\ldots)$, 
    $(\ldots,\underline{0},0,1,0\ldots)$, 
    $(\ldots,\underline{0},0,0, 1,0\ldots)$,  and $(\ldots,\underline{0},0,0, 0, 1,0\ldots)$, where we use the underline to denote  $i_0$.    
    Each solution is localized at a local minima of the quasiperiodic potential. The lack of translational invariance implies that all displayed solutions are distinct, despite their codes differing only in the location of the symbol ``1'' relative to $i_0=\underline{0}$. The coding approach can also be applied to more complex solutions:  in Fig.~\ref{fig:solitons}(b), we show the  solutions with the codes $(\ldots,\underline{0},1,1,0,0,\ldots)$ and $(\ldots,\underline{0},0, 0, 1,1,0\ldots)$.

    Let us briefly discuss the range of the parameters where the established coding can be applied. First, we recall that the validity of Hypotheses~\ref{hypothesis:island-set} and \ref{hypothesis:strips-mapping} for $\delta=0$ imply that the same hypotheses remain valid for any $\delta\in[0;2\pi)$, see the remark in Sec.~\ref{sec:SymbDyn}.
    In addition, we fulfilled the same numerical check for  $\mu = 1$, $A_1 = 3$, and various values of $A_2$. Our computations show that Hypotheses \ref{hypothesis:island-set} and \ref{hypothesis:strips-mapping} hold if  $0 \leq |A_2| \leq 2.5$.
    Note that at $A_2=0$ the potential becomes periodic, and the coding approach reduces to that introduced  in Ref.~\cite{AA13}. 

    \ga{It is instructive to compare our results with those obtained using rational approximations of the golden ratio $\theta$, a common technique for quasiperiodic problems \cite{Zilber2021,Modugno10,Diener,M09,LiLiS17,PKZ22,ZA24,Konotop24}. We applied our method to Eq.~\eqref{eq:general-form-equation} with $\theta$ replaced by the rational approximations $21/13$ and $55/34$, where the numerator and denominator of each fraction are successive Fibonacci numbers.  In both cases, the potential $V(x)$ becomes periodic, with periods   $13 \pi$  and $34\pi$, respectively.  For both approximations, we observed no appreciable difference from the results for the original quasiperiodic potential.  Specifically, with identical parameters ($\mu$, $A_1$, $A_2$), three islands appear in each $\Delta$-slice ($\Delta\in[0,2\pi)$), corresponding to three donuts in the full   space $\mathcal{L}$.   Analysis of the strips $V_{i,j}$ shows that Hypothesis~\ref{hypothesis:strips-mapping} also holds. Therefore,  there exists the one-to-one correspondence between all bi-infinite sequences over the  three-symbol alphabet $\mathcal{A}_3=\{-1,0,1\}$ and the regular,  real-valued solutions of Eq.~\eqref{eq:general-form-equation}   where $\theta$ is replaced by its rational approximation. This implies, in particular, that
    $\mu=1$ must lie within the spectral gaps of both approximate periodic potentials. If this were not the case, the behavior of ${\mathcal P}$ in vicinity of the origin $(0,0)$ would not be hyperbolic, and the matrix ${\mathcal D}{\mathcal P}_{(0,0)}$  would fail to satisfy conditions (H-1)–(H-2). However, one cannot generally exclude the situation where a given $\mu$ falls within a spectral gap for one approximate potential but lies in an allowed band for the other.}     
    
   To demonstrate that the  one-to-one correspondence established  in  Example~\ref{exa:example-1} is not  trivial and does not hold by default, below we provide two examples where one of  the hypotheses fails, and consequently, the  coding procedure is not directly applicable.
       
    \begin{figure*}[ht]
		\centering
        \includegraphics[width=.9\linewidth]{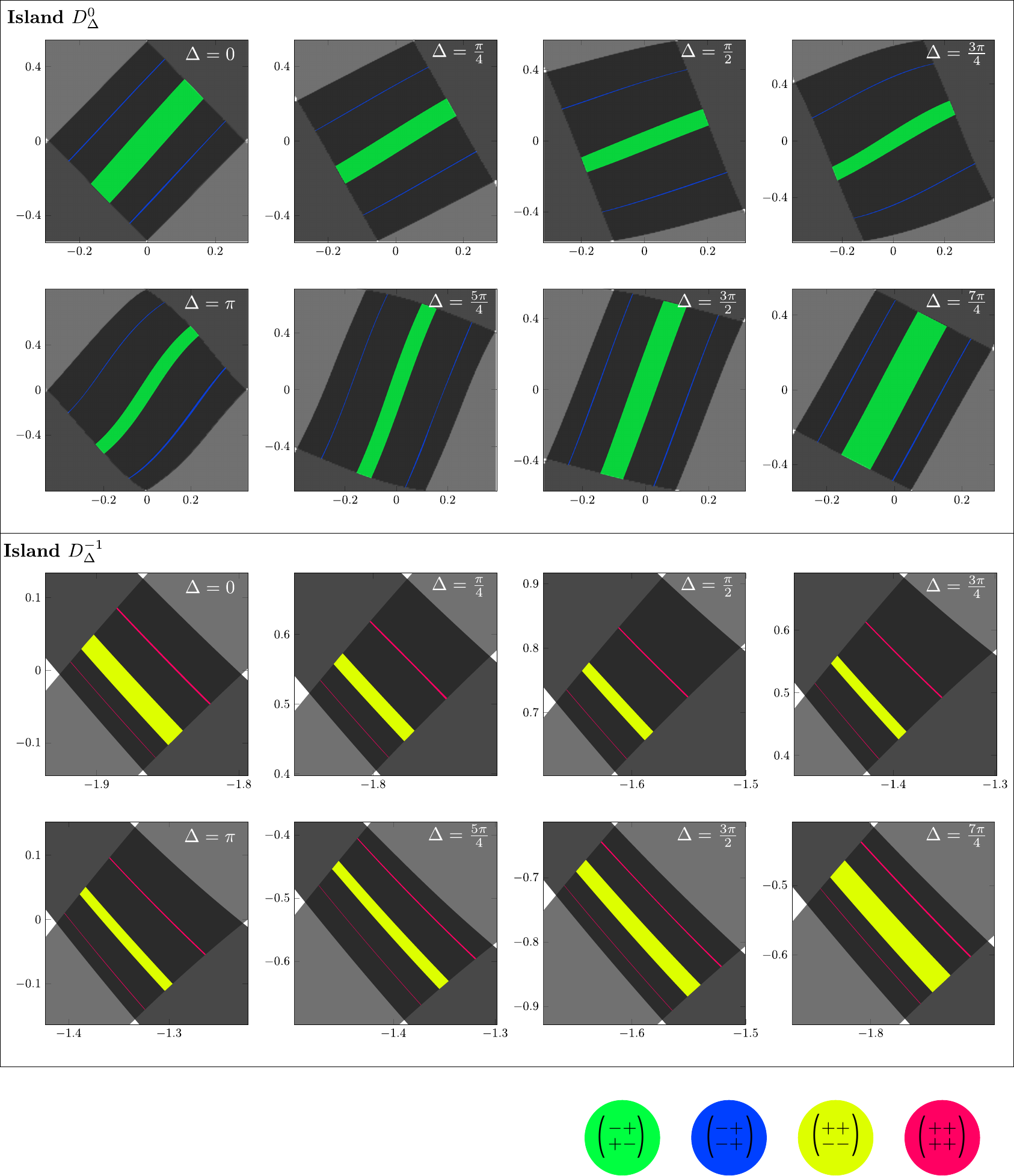}
		\caption{Numerical verification of Hypothesis~\ref{hypothesis:strips-mapping} for Example \ref{exa:example-1}. Each panel shows $v$-strips $V_{i,j}(\Delta)$, $i,j\in\{-1,0,1\}$ in the island $D_\Delta^0$ (upper panel) and $D_\Delta^{-1}$ (lower panel), for  $\Delta=k\pi/4$, $k=0,1,\ldots,7$. The   configurations of signs in the linearization matrices ${\mathcal D}{\mathcal P}_{\bf p}$ are represented  by different colors, see the color legend under the main panels.}
	    \label{fig:example-1-diagram-islands-grids}
	\end{figure*}

    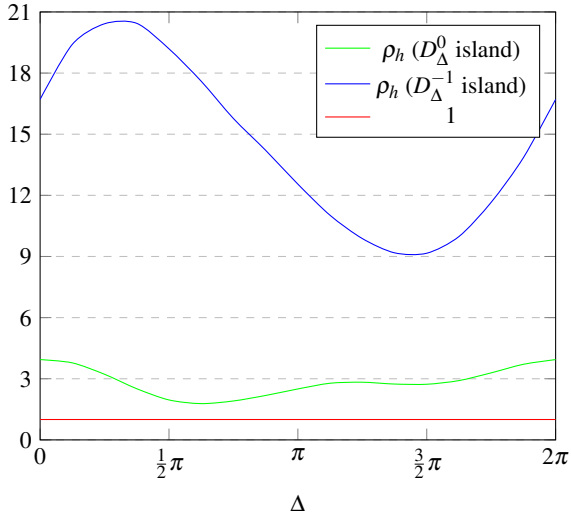
\begin{figure}[ht]
        \centering
        \begin{tikzpicture}
            \begin{axis}[
                title = {},
                xlabel = {$\Delta$},
                ylabel = {},
                xmin = 0,
                xmax = 360,
                ymin = 0,
                ymax = 21,
                xtick = {0, 90, 180, 270, 360},
                xticklabels={$0$, $\frac{1}{2} \pi$, $\pi$, $\frac{3}{2} \pi$, $2 \pi$},
                ytick = {0, 3, ..., 21},
                legend pos = north east,
                ymajorgrids = true,
                grid style = dashed,
            ]
                
            \addplot [
                    color = green,
                    smooth,
                ] coordinates {
                    (0, 3.94)
                    (22.5, 3.78)
                    (45, 3.22)
                    (67.5, 2.51)
                    (90, 1.96)
                    (112.5, 1.78)
                    (135, 1.91)
                    (157.5, 2.18)
                    (180, 2.5)
                    (202.5, 2.78)
                    (225, 2.83)
                    (247.5, 2.74)
                    (270, 2.73)
                    (292.5, 2.91)
                    (315, 3.29)
                    (337.5, 3.71)
                    (360, 3.94)
                };
            \addlegendentry{$\rho_h$ ($D_{\Delta}^{0}$ island)}
            \addplot [
                    color = blue,
                    smooth,
                ] coordinates {
                    (0, 16.71)
                    (22.5, 19.41)
                    (45, 20.42)
                    (67.5, 20.44)
                    (90, 19.19)
                    (112.5, 17.6)
                    (135, 15.77)
                    (157.5, 14.21)
                    (180, 12.55)
                    (202.5, 11.02)
                    (225, 9.88)
                    (247.5, 9.18)
                    (270, 9.16)
                    (292.5, 9.98)
                    (315, 11.66)
                    (337.5, 13.85)
                    (360, 16.71)
                };
            \addlegendentry{$\rho_h$ ($D_{\Delta}^{-1}$ island)}
            \addplot [
                color = red
                ]
                coordinates {
                    (0, 1)
                    (360, 1)
                };
            \addlegendentry{$1$}
            \end{axis}
        \end{tikzpicture}
        
        \caption{Values of $\rho_h$ for $D_\Delta^{-1}$ and $D_\Delta^0$ for Example~\ref{exa:example-1}. The values of $\rho_h$ for $D_\Delta^{1}$ are the same as for  $D_\Delta^{-1}$ since these islands are symmetric to each other with respect to the origin.}
        \label{fig:example-1-island-1-mu-graph}
    \end{figure}

 \begin{figure}
		\centering
        \includegraphics[width=\linewidth]{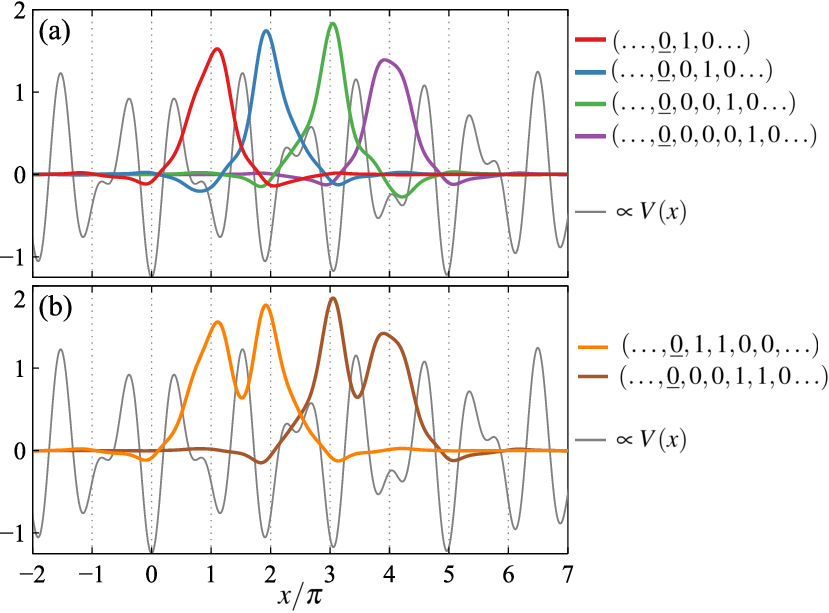}
		\caption{Example~\ref{exa:example-1}: solutions $u(x)$ with codes that contain exactly one nonzero symbol (a) and exactly two nonzero symbols (b). Thin gray lines plot the quasiperiodic lattice (up to a scaling factor).}
	    \label{fig:solitons}
	\end{figure}

\end{exa}

\begin{exa}
    \label{exa:example-2}

    If in Eq.~\eqref{eq:general-form-equation} we take $\mu=0.4$, while keeping $A_1 = 3$,  $A_2 = 2$ and $\delta=0$ as in Example~\ref{exa:example-1}, the situation changes qualitatively.
    Figure~\ref{fig:example-2} shows the sets ${\mathscr U}^+_{\Delta}$ (dark gray), ${\mathscr U}^-_{\Delta}$ (light gray) for $\Delta=0$ and $\pi$. For  $\Delta=\pi$, the  intersection ${\mathscr U}_{\Delta}$ (black) can not be represented as an union of disjoint islands. Consequently, the set ${\mathscr U}$ is not a donut set, and the conditions of Hypothesis~\ref{hypothesis:island-set} are not satisfied. Thus, the coding procedure does not apply in this case.
        

    \begin{figure}[ht]
		\centering
        \includegraphics[width=\linewidth]{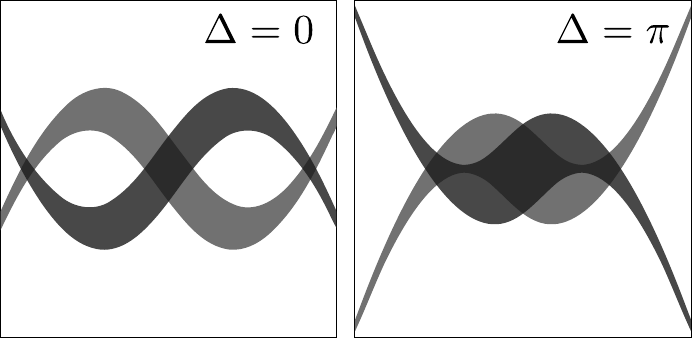}
		\caption{Example~\ref{exa:example-2}: for $\Delta=\pi$ island are not separated and the conditions of Hypothesis~\ref{hypothesis:island-set} are not satisfied.  Each panel shows the region $(u, u') \in [-2; 2] \times [-2; 2]$.}
	    \label{fig:example-2}
	\end{figure}
\end{exa}

\begin{exa}
    \label{exa:example-3}

    For the parameters of Example~\ref{exa:example-1} but with $\mu=1.8$ instead of $\mu=1$, the island structure persists in every $\Delta$-slice which indicates that Hypothesis~\ref{hypothesis:island-set} holds. However, checking the   entries in the matrices ${\mathcal D}{\mathcal P}_{\bf p}$ reveals that their signs do not satisfy the conditions of Theorem~\ref{thm:h-strips-mapping} (see Fig.~\ref{fig:example-3}). Therefore, we cannot claim that Hypothesis~\ref{hypothesis:strips-mapping} also holds, and our coding procedure is again inapplicable.

    \begin{figure}[ht]
		\centering
        \includegraphics[width=\linewidth]{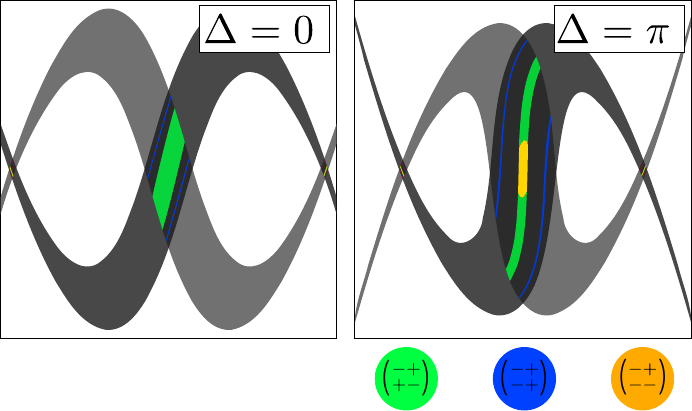}
		\caption{Example~\ref{exa:example-3}: the signs of linearization matrices do not meet Theorem~\ref{thm:h-strips-mapping} conditions. Each panel shows the region $(u, u') \in [-2.3; 2.3] \times [-2; 2]$.}
	    \label{fig:example-3}
	\end{figure}
\end{exa}

    \section{Alternative interpretation of the coding approach} \label{Sec:Alt_Int}

\ga{Here we provide a heuristic interpretation of our mathematical results on coding of nonlinear states in a quasiperiodic potential. This interpretation builds upon known results regarding the solutions of the equation
\begin{gather}
u_{xx}+(\mu-V(x))u-u^3
=0.\label{StatGPEq_u3}
\end{gather}
}
\ga{Let $\mu$ be fixed. We split the explanation into the following steps.
}

\begin{figure}
		\centering
        \includegraphics[width=\linewidth]{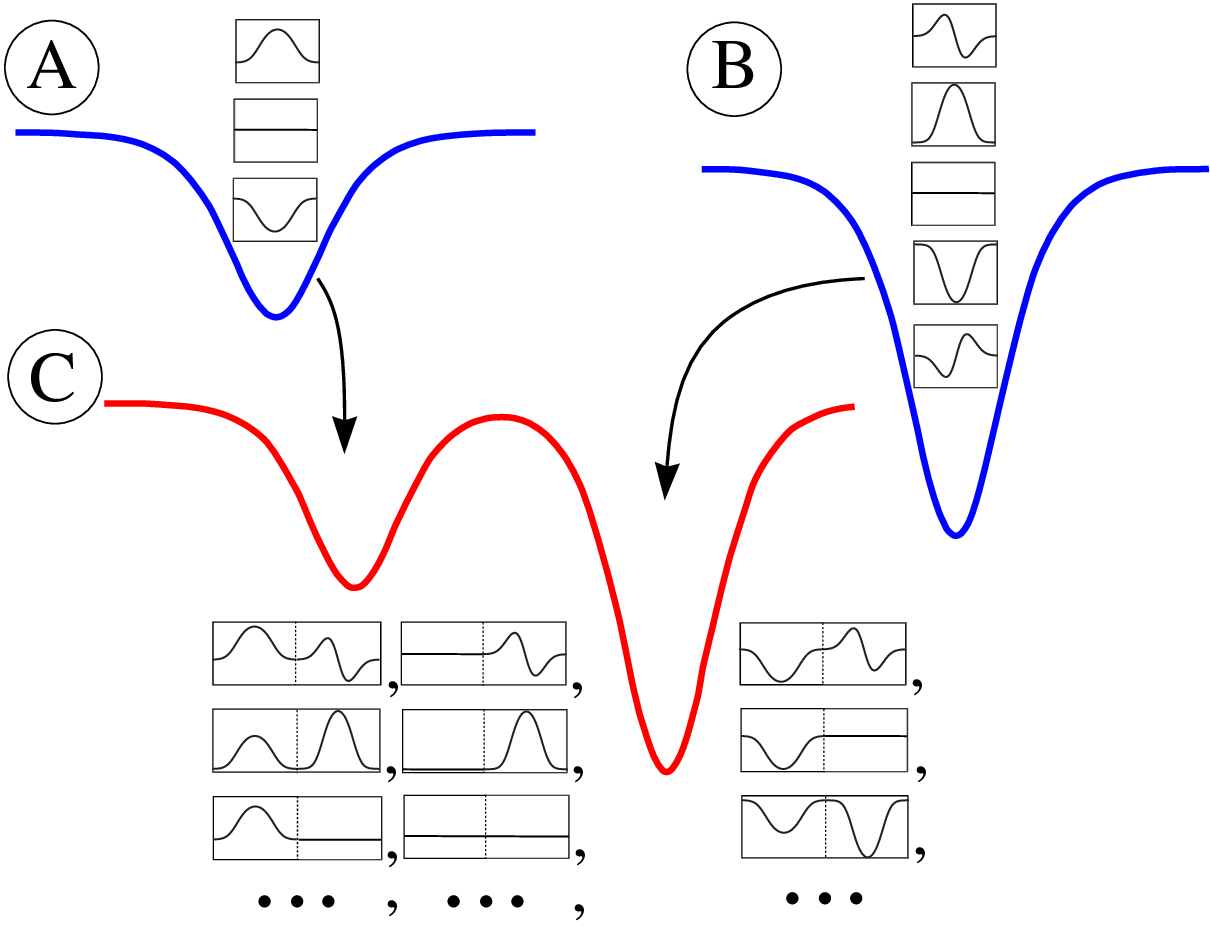}
		\caption{Nonlinear states in a double-well potential (C), constructed as bound states of the individual well solutions (A and B).}
	    \label{fig:Double_Well}
	\end{figure}

\ga{1. Let $V(x)$ be a single-well potential of finite depth (see Fig.~\ref{fig:Double_Well}A,B). Consider localized solutions of (\ref{StatGPEq_u3}) such that 
\begin{gather*}
    u(x)\to 0,\quad x\to\pm\infty.
\end{gather*}
Evidently, $u(x)\equiv 0$ is a solution of Eq.~(\ref{StatGPEq_u3}). Also, if $u(x)$ is a solution of Eq.~(\ref{StatGPEq_u3}) then $-u(x)$ is also a solution of this equation. The total number of localized solutions is finite, odd and  determined by the width and the depth of the potential $V(x)$. 
We note that arguments for a finite number of nonlinear states under the repulsive interactions are similar to those presented in Ref.~\cite{AZ07}. 
}

\ga{2. Let $V(x)$ be a double-well potential of finite depth (see Fig.~\ref{fig:Double_Well}C). Then, under certain restrictions, the  solutions of Eq.~(\ref{StatGPEq_u3}) can be   approximated by bound states composed of the solutions confined to  each individual well.  This statement is evident for well-separated potential wells but also holds when the wells are close.  
}

\ga{3. Let $V(x)$ be a periodic potential. It can be viewed as an infinite chain of identical, equally spaced wells of finite depth. The states localized within each elementary well can then serve as "building blocks" for constructing solutions of the full potential $V(x)$. This approach was employed in Refs.\cite{Comp_Rel01,Comp_Rel02}, where the authors introduced a so-called "composition relation" linking the states defined on individual periods to   solutions of the full equation. A rigorous justification for this treatment, in the form of a coding procedure, was provided in \cite{AA13, AL24}. In these works, conditions were established that guarantee   {\it any} combination of elementary cell states yields a valid nonlinear state of the periodic potential.
}

\ga{4. Now assume that $V(x)$ is a quasiperiodic  potential, which can be  viewed  as an infinite chain of finite depth potential wells. These wells are not fully identical and cannot be regarded as strictly equally spaced (see Fig.~\ref{fig:solitons}, gray line). From this viewpoint, the main result of the present paper consists in  the conditions that guarantee the applicability of the ``composition relation''  for such a  quasiperiodic ---   specifically, bichromatic --- potential. 
} 

    \begin{figure}
		\centering
        \includegraphics[width=\linewidth]{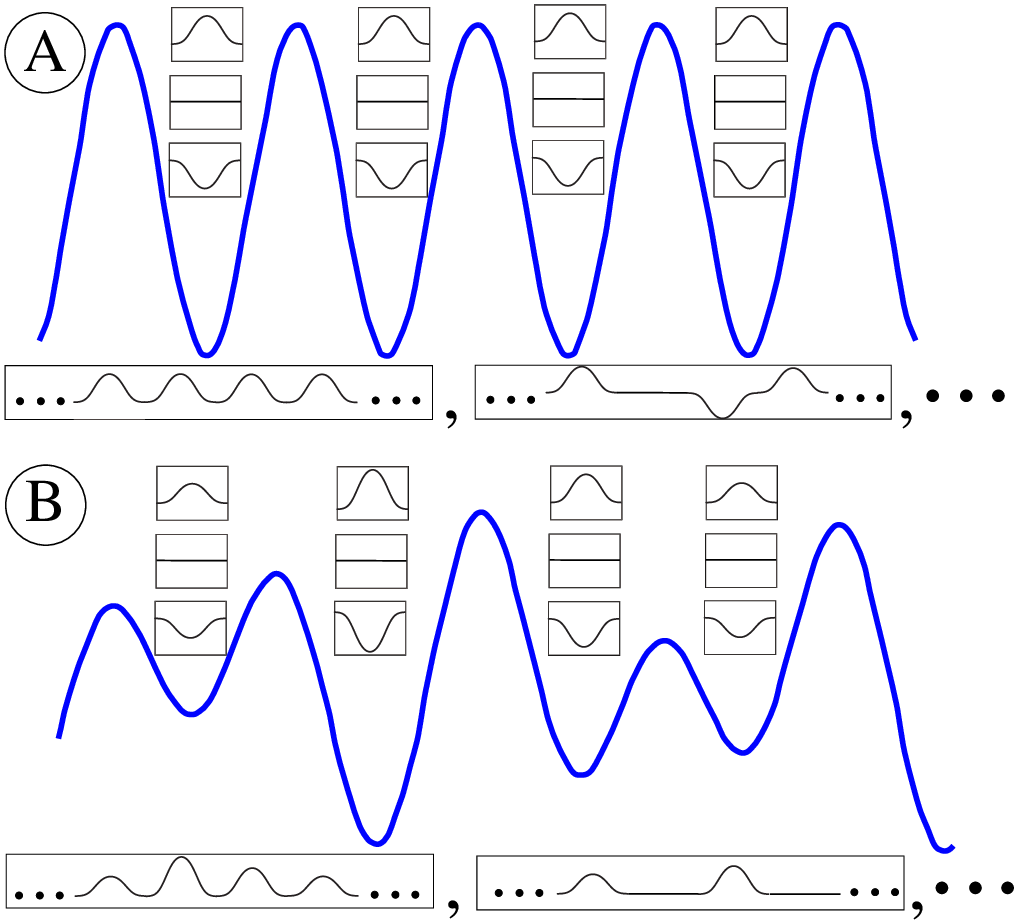}
		\caption{Nonlinear states in (A) periodic and  (B) quasiperiodic   potentials, constructed as bound states of the individual well solutions.}
	    \label{fig:Per_QuasiPer}
	\end{figure}

\section{Conclusion} \label{Sec:Concl}

In this paper, we have developed a coding approach for stationary states of the one-dimensional Gross-Pitaevskii equation (GPE) with   two  incommensurate spatial frequencies. In the application to the theory of Bose-Einstein condensates, the incommensurate terms can originate from different physical mechanisms. The first mechanism is a superposition of two laser beams that form a quasiperiodic optical lattice. The second mechanism arises from a periodic modulation of the scattering length with a frequency that is incommensurate with that of the optical trap. The third possibility corresponds to the quasiperiodic scattering length.

We have demonstrated that, under certain conditions, the stationary states can be described using bi-infinite sequences of symbols from a finite alphabet.  Our approach leverages the fact that, when the condensate is dominated by the  repulsive interactions, a large subset of solutions to the Cauchy problem for the corresponding ordinary differential equation tend to infinity at some finite point on the real axis. 
The coding approach has been illustrated with a numerical example for the GPE with a quasiperiodic lattice.  


Finally, we outline some problems amenable to the method developed in this work and suggest directions for its future development. A primary direction is the study of quasiperiodic versions of the GPE with a spatially modulated scattering length. This modulation could be periodic (with a frequency incommensurate with the optical trap) or quasiperiodic (resulting from the interplay of two incommensurate modulations). 
These studies should be followed by the analysis of stability of these nonlinear states keeping in mind possible relation between the codes of solutions and their stability.

The method can be further developed in several directions. First, it is promising to weaken the assumptions of Hypotheses \ref{hypothesis:island-set} and \ref{hypothesis:strips-mapping}. This may be achieved by considering the iterated maps ${\mathcal P}^n$ and ${\mathcal P}^{-n}$ for $n \geq 2$ instead of ${\mathcal P}$ and ${\mathcal P}^{-1}$. Numerical simulations suggest that using iterated maps leads to a contraction of the $h$- and $v$-strips, which could potentially extend the range of parameters for which the procedure is applicable. Second, in its current form, Theorems \ref{thm:h-strips-mapping} and \ref{thm:v-strips-mapping} are formulated in terms of the sign patterns of the entries of the Jacobian ${\mathcal D}{\mathcal P}$, which determine the mapping of quadrants. These theorems can likely be strengthened by considering arbitrary pairs of non-intersecting cones, rather than being restricted to quadrants. It would be valuable to extend the approach to cases with more complex nonlinearities and non-real-valued solutions.

\ga{A possible ramification of our approach could involve an application to Thouless pumping \cite{Zilber2021,pump} of  linear wave packets and solitons in periodic and quasiperiodic bichromatic lattices. In these studies (e.g.,  Refs.\cite{soltransport1,soltransport2}),  topological transport is achieved   by adiabatically increasing the phase shift $\Delta$ between the constituent  sublattices. In the context of the present work,  one topological pumping  cycle   would correspond to an adiabatic circuit   around the donut as  $\Delta$ traverses  the  full  space $\S$.}




\section*{Acknowledgment} \label{Sec:ask}


The research of D.A.Z. was supported by Russian Science Foundation, Grant No. 25-12-00119, Ref.\cite{RNF}.

\section*{Data Availability} \label{Sec:Data}
The data that supports the findings of this study are available from the corresponding author upon reasonable request.
\nocite{*}

\end{document}